\pdfoutput=1
\documentclass[12pt]{article}
\usepackage[english]{babel}
\usepackage{algorithm}
\usepackage{algorithmic}
\usepackage{amsfonts}
\usepackage{amsmath}
\usepackage{amssymb}
\usepackage{bbm}
\usepackage{blindtext}
\usepackage{booktabs}
\usepackage{color}
\usepackage{geometry}
\usepackage{hyperref}
\usepackage{tabularx}
\usepackage{theorem}
\usepackage{times}
\usepackage{url}
\usepackage{graphicx}
\usepackage{setspace}
\newcommand{\diag}{\text{diag}}
\newcommand{\eg}{\textit{e.g.}}
\newcommand{\etal}{\textit{et~al}.}

\newcommand{\va}{\mathbf{a}}
\newcommand{\vb}{\mathbf{b}}

\newcommand{\vw}{\mathbf{w}}
\newcommand{\vx}{\mathbf{x}}
\newcommand{\vy}{\mathbf{y}}
\newcommand{\vz}{\mathbf{z}}

\newcommand{\zero}{\mathbf{0}}

\renewcommand{\d}{\text{d}}

\newcommand{\mA}{\mathbf{A}}
\newcommand{\mB}{\mathbf{B}}

\newcommand{\mD}{\mathbf{D}}

\newcommand{\mM}{\mathbf{M}}

\newcommand{\mX}{\mathbf{X}}

\newcommand{\eps}{\epsilon}
\newcommand{\valpha}{\text{\boldmath{$\alpha$}}}
\newcommand{\vbeta}{\text{\boldmath{$\beta$}}}

\newcommand{\veps}{\text{\boldmath{$\epsilon$}}}

\newcommand{\vgamma}{\text{\boldmath{$\gamma$}}}

\newcommand{\vmu}{\text{\boldmath{$\mu$}}}

\newcommand{\vtheta}{\text{\boldmath{$\theta$}}}

\newcommand{\mDelta}{\mathbf{\Delta}}

\newcommand{\mSigma}{\mathbf{\Sigma}}

\newcommand{\B}{\mathcal{B}}

\newcommand{\DP}{\mathcal{DP}}

\newcommand{\G}{\mathcal{G}}
\newcommand{\IG}{\mathcal{IG}}

\newcommand{\N}{\mathcal{N}}

\newcommand{\U}{\mathcal{U}}

\newcommand{\dr}{\textit{dr}}
\newcommand{\mine}{\textit{mine}}
\newcommand{\agri}{\textit{agri}}
\newcommand{\ind}{\textit{ind}}
\newcommand{\enrol}{\textit{enrol}}
\newcommand{\child}{\textit{child}}
\newcommand{\old}{\textit{old}}
\newcommand{\lnarea}{\textit{lnarea}}
\newcommand{\lndoc}{\textit{lndoc}}
\newcommand{\dum}{D(t)}
\newcommand{\yaf}{\text{1930}}
\newcommand{\ybf}{\text{1925}}

\newcommand{\jasa}{\textit{Journal of the American Statistical Association}}
\newcommand{\csda}{\textit{Computational Statistics \& Data Analysis}}
\newcommand{\joe}{\textit{Journal of Econometrics}}
\newcommand{\biometrika}{\textit{Biometrika}}
\newcommand{\econometrica}{\textit{Econometrica}}
\newcommand{\sac}{\textit{Statistics and Computing}}
\newcommand{\jbes}{\textit{Journal of Business and Economic Statistics}}
\newcommand{\jae}{\textit{Journal of Applied Econometrics}}

\title{\bf
{Bayesian Nonparametric Instrumental Variable Regression Approach to Quantile Inference}}
{
\author{Genya Kobayashi\\
Faculty of Law, Politics \& Economics, Chiba University\\
\texttt{gkobayashi@chiba-u.jp}
\and
Kota Ogasawara\\
Department of Industrial Engineering and Economics, Tokyo Institute of Technology\\
\texttt{ogasawara.k.ab@m.titech.ac.jp}
}}

\begin{document}

\maketitle

\begin{abstract}
This study extends the Bayesian nonparametric instrumental variable regression model to determine the structural effects of covariates on the conditional quantile of the response variable.
The error distribution is nonparametrically modelled using a Dirichlet mixture of bivariate normal distributions.
The mean functions include the smooth effects of the covariates represented using the spline functions in an additive manner.
The conditional variance of the second-stage error is also modelled using the spline functions such that it varies smoothly with the covariates.
Accordingly, the proposed model allows for considerable flexibility in the shape of the quantile function while correcting for an endogeneity effect.
The posterior inference for the proposed model is based on the Markov chain Monte Carlo method that requires no Metropolis--Hastings update.
The approach is demonstrated using simulated and real data on the death rate in Japan during the inter-war period.

\noindent
\textbf{keywords}: Dirichlet process mixture; Historical data; Instrumental variable; Markov chain Monte Carlo; Penalised splines

\end{abstract}

\newpage
\section{Introduction}
A set of quantiles provides a more complete description of a distribution than the mean. Similarly, exploring the relationship between the quantiles of the conditional distribution of the response variable and a set of covariates offers key insight into the structure of the data at hand.
As an important approach to quantile inference, the quantile regression has received substantial attention since Koenker and Bassett's~(1978) seminal work.
There is a vast literature on the theory of quantile regression; for an overview, see Koenker~(2005), Yu~\etal~(2003), and Buchinsky~(1998).

Studies on Bayesian quantile regression particularly increased following Yu and Moyeed~(2001), who introduced a likelihood function on the basis of asymmetric Laplace distribution and performed a posterior estimation using the Markov chain Monte Carlo (MCMC) method.
A wide variety of quantile regression models based on the asymmetric Laplace distribution have been considered in the literature (\eg~Kozumi and Kobayashi,~2011; Yue and Rue,~2011; Lum and Gelfand,~2012; Neelon~\etal,~2014).
Moreover, Kottas and Krnjaji\'c~(2009) and Reich~\etal's~ (2010) semiparametric approaches aimed at constructing a more flexible error distribution using quantile restriction.
As an alternative to modelling error distributions with quantile restrictions, recent studies have directly considered modelling a conditional distribution or quantile function of the response variable (\eg~Taddy and Kottas,~2010; Kadane and Tokdar,~2012; Reich~\etal,~2011; Jang and Wang,~2015).

As in the case of conventional mean regression, a standard quantile regression estimator is known to be biased when an endogenous variable is included.
The problem of endogeneity in the quantile regression framework was recognised as early as in the 1980s (Amemiya,~1982; Powell,~1983).
Since then, there has been a growing interest in the inference for quantile regression models with endogenous variables and quantile treatment effects in the context of the frequentist approach (\eg\ Abadie~\etal,~2002; Kim and Muller,~2004; Ma and Koenker,~2005; Chernozhukov and Hansen,~2005; 2006; 2008; Lee,~2007; Horowitz and Lee,~2007; Blundell and Powell,~2007; Chernozhukov~\etal,~2015).

There exist numerous studies that adopt the Bayesian framework to examine mean regression models with an endogenous variable from a theoretical and computational viewpoint (\eg\ Hoogerheide~\etal,~2007a; 2007b; Conley~\etal,~2008; Lopes and Polson,~2014; Wiesenfarth~\etal,~2014).
However, despite the growing interest in and demand for Bayesian quantile inference, the literature on the Bayesian approach to quantile inference in the presence of an endogenous variable remains particularly sparse.
Lancaster and Jun~(2010) adopted the moment conditions adopted in Chernozhukov and Hansen~(2006) for the empirical likelihood.
Drawing on Lee~(2007), Kobayashi~(2016) and Ogasawara and Kobayashi~(2015) employed the control function approach.
They considered the use of parametric and semiparametric first-stage error distributions, whose quantile is restricted to zero for some quantile level, and employed the asymmetric Laplace distribution in the second stage, as in the standard Bayesian quantile regression.
Kobayashi's~ (2016) simulation study showed that while this approach works well in many situations, it may provide biased estimates owing to the restrictive specification of the first-stage error distribution.
Furthermore, both approaches assume a simple linear model for the conditional quantile and monotonicity (non-crossing) of the estimated quantiles is not necessarily guaranteed.

Given the preceding discussion, this study aims to provide a flexible Bayesian approach to determine the structural effects of covariates on the conditional quantile of the response variable in the presence of an endogenous variable.
To this end, we extend the Bayesian nonparametric instrumental variable regression model proposed by Wiesenfarth~\etal~(2014) and Conley~\etal~(2008) and consider a location-scale specification.
In particular, the joint error distribution of the instrumental variable regression model is modelled using the Dirichlet process mixture of bivariate normals to capture the non-normality of the error distribution.
To capture the nonlinear effect of the covariate on the response variable, the mean functions are nonparametrically modelled using the spline functions.
In addition, we also nonparametrically model the conditional variance of the second-stage regression using spline functions such that the conditional variance varies smoothly with covariates.
Unlike the quantile curves implied from the Bayesian nonparametric instrumental variable regression model, those implied from the proposed model are not necessarily parallel to each other since the conditional variance differs by covariate value.
Therefore, the estimated quantile curves allow significant flexibility in estimating the structural effect of the covariate of interest.
In addition, since the conditional quantiles are computed from the nonparametrically estimated error distribution, the monotonicity of the quantile curves is guaranteed.
The proposed model is estimated using the MCMC method that includes no Metropolis--Hastings (MH) update.
The MCMC method is applied to sample from the posterior distributions of the conditional quantiles.

The remainder of this paper is organised as follows.
The proposed model is introduced and the choice of the prior distributions is discussed in Section~\ref{sec:model}.
The MCMC method for the posterior inference on the conditional quantiles is developed in Section~\ref{sec:mcmc}.
The proposed approach is illustrated using the simulated data in Section~\ref{sec:sim} and real data in Section~\ref{sec:real}, where the effect of doctors on the death rate in Japan during the inter-war period is studied.
Finally, we conclude in Section~\ref{sec:conc}.

\section{Proposed Approach}\label{sec:model}
\subsection{Bayesian Instrumental Variable Regression Model}
The following parametric Bayesian instrumental variable regression model provides the starting point of the model proposed in this study:
\begin{equation}\label{eqn:iv}
\begin{split}
d &=\gamma_0+z\gamma_1+ \eps_{1},\\
y &=\beta_0+d\delta+\eps_{2},
\end{split}
\end{equation}
where $y$ is the response variable, $d$ is the endogenous variable, $z$ is the instrumental variable, $\gamma_0$ and $\beta_0$ are the intercepts, $\gamma_1$ and $\delta$ are the coefficient parameters, and $\veps=(\eps_{1},\eps_{2})'$ is the error term following $\N(\zero,\mSigma)$, where
$$
\mSigma = \left[
\begin{array}{cc}
\sigma_{11}&\sigma_{12}\\
\sigma_{12}&\sigma_{22}\\
\end{array}
\right].
$$
The correlation coefficient $\rho=\sigma_{12}/\sqrt{\sigma_{11}\sigma_{22}}$ represents the degree of endogeneity.

The normality assumption for $\veps$ can be restrictive.
Outliers, some form of model misspecification, and heterogeneity can also induce non-normality in the error term. 
To relax the normality assumption, Conley~\etal~(2008) extended the model \eqref{eqn:iv} by introducing the Dirichlet process mixture of bivariate normal distributions to nonparametrically construct the joint distribution of $\veps$.
More recently, Wiesenfarth~\etal~(2014) further extended the model of Conley~\etal~(2008) by considering
\begin{equation}\label{eqn:niv}
\begin{split}
d &=\gamma_0 +f(z) + \eps_1,\\
y &=\beta_0+ g(d)+ \eps_2,
\end{split}
\end{equation}
where $\gamma_0$ and $\beta_0$ are intercepts and $f(z)$ and $g(d)$ represent the nonlinear effects of the instrument and endogenous variables, respectively.
The assumptions required for identification are denoted by $E[\eps_1|z]=0$ and $E[\eps_2|\eps_1,z]=E[\eps_2|\eps_1]$.
The smooth functions are approximated using the spline functions and are represented as the linear combinations of the basis functions.
The extensive simulation in these studies revealed that the nonparametric modelling approach is more robust against non-normality and outliers and performs better than existing approaches, including two- and three-stage least squares, a model under the normality assumption, and the control function approach with the generalised cross-validation procedure for the choice of a smoothing parameter.

From the viewpoint of quantile inference, Conley~\etal~(2008) and Wiesenfarth~\etal~'s(2014) modelling approach could be used to infer the structural effect of the covariate on the conditional quantile.
However, although their Dirichlet process mixture approach to the joint error distribution introduces heteroskedasticity over observations, the estimated quantile curves are parallel to each other since the conditional variance does not vary by covariate.
An interesting feature of quantile regression is that its coefficients can vary by quantile.
For example, if the data are generated from $y=\vx'\vbeta+(\vx'\vgamma)\eps$, where $\eps$ is the error term and $\vx'\vgamma$ represents heteroskedasticity, the $p$-th quantile regression estimator corresponds to $\vbeta+\vgamma Q_\eps(p)$, where $Q_\eps(\cdot)$ is the quantile function of $\eps$ (He,~1997; Reich~\etal,~2010).
Therefore, the applicability of Conley~\etal~(2008) and Wiesenfarth~\etal~(2014) to quantile inference would be limited to cases in which error distribution may exhibit heteroskedasticity but the variance does not depend on the covariates.

\subsection{Nonparametric Model for Quantile Inference}
To allow for inference on the conditional quantile while correcting for an endogeneity effect, we extend \eqref{eqn:niv} and consider a location-scale model for more flexibility in the quantile curve.
Furthermore, to introduce the model for the conditional variance and prior distribution in a more flexible manner, the error term is rewritten following Lopes and Polson~(2014).
Specifically, this study considers the model given by
\begin{equation}\label{eqn:mod}
\begin{split}
d_i &=\gamma_0 +f(z_i) + v_i,\quad v_i\sim\N(\mu_{1i},\tau_i)\\
y_i &=\beta_0+ g(d_i)+ \eta_i(v_i-\mu_{1i}) +  e_i, \quad e_i\sim\N(\mu_{2i},s_i\sigma_i)\\
s_i &=\exp\left\{\alpha_0+ h(d_i) \right\},
\end{split}
\end{equation}
for $i=1,\dots,n$, where
$\alpha_0$ is the intercept and $h$ is the smooth function of $d$.
For identifiability, the functions $f(\cdot)$, $g(\cdot)$, and $h(\cdot)$ are centred around zero.
This formulation implies that the components of the covariance matrix $\mSigma_i$ for $(\eps_{1i},\eps_{2i})'$ is recovered through $\sigma_{11i}=\tau_i$, $\sigma_{12i}=\eta_i\sigma_{11_i}$, and $\sigma_{22i}=s_i\sigma_i+\sigma_{12i}^2/\sigma_{11i}$.
Although each equation in \eqref{eqn:mod} includes only one nonlinear effect for notational simplicity, the model can include more functions of covariates in an additive manner.

As in Conley~\etal~(2008) and Wiesenfarth~\etal~(2014), the error distribution is nonparametrically modelled using the Dirichlet process mixture, but the prior distribution is introduced in a different manner.
We introduce the Dirichlet process prior for the distribution of $\vtheta_i=(\mu_{1i},\mu_{2i},\eta_i,\tau_i,\sigma_i)$:
\begin{equation*}
\begin{split}
\vtheta_i&\sim G,\quad i=1,\dots,n,\\
G&\sim\DP(a,G_0),
\end{split}
\end{equation*}
where $\DP(a,G_0)$ denotes the Dirichlet process with the precision parameter $a$ and base measure $G_0$.
Since our model contains the global intercepts $\gamma_0$ and $\beta_0$, to achieve identifiability, we employ the priors $\mu_1$ and $\mu_2$ such that $\mu_1$ and $\mu_2$ have mean zero apriori and force $\sum_{i=1}^n\mu_{1i}=\sum_{i=1}^n\mu_{2i}=0$ to ensure they have zero mean posteriori.
In addition, since $s_i$ includes the intercept $\alpha_0$, we force $\sum_{i=1}^n\log\sigma_i=0$ posteriori as in Sarkar~\etal~(2014).
As discussed in Wiesenfarth~\etal~(2014), this nonparametric modelling approach to error distribution allows different degrees of endogeneity correction over observations and downweights outliers in the error term.
The degree of downweighting is controlled through the precision parameter of the Dirichlet process and the parameters for variance components.
The representation for the error term in \eqref{eqn:mod} allows for a more flexible choice of base measure and hyperpriors than that in Conley~\etal~(2008) and Wiesenfarth~\etal~(2014), where the inverse Wishart distribution for $\mSigma_i$ was employed (Lopes and Polson,~2014).
In this study, we employ the following independent distributions for $G_0$: $\mu_1\sim\N(\vartheta_{1},M_{10})$, $\mu_2\sim\N(\vartheta_{2},M_{20})$, $\eta\sim\N(\vartheta_e,\varphi_e)$, $\tau\sim\IG(t_1,t_2)$, and $\sigma\sim\IG( s_1,s_2)$.
The choice of the hyperparameters and hyperpriors is discussed in Section~\ref{sec:prior}.

Each smooth function of covariate is approximated using a linear combination of B-spline functions in the study.
For $f(\cdot)$, for example, given the number of interior knots $K_{f}$ and degree $m_f$, let us denote the set of knots by
$\kappa_{f}=\left\{\right.
\kappa_{f,1}=\cdots=\kappa_{f,m_f+1}<\kappa_{z,m_z+2}<\cdots<\kappa_{f,m_f+K_{f}+1}=\cdots=\kappa_{f,2m_f+K_{f}+1}
\left.\right\}$.
Using these knots, the B-spline bases of degree $m_{f}$ denoted by $\vb_{K_{f},m_f}(z)=(b_{K_{f},1}(z),\dots,b_{K_{f},K_{f}+m_f}(z))$ can be calculated through a simple recursion of de~Boor~(2000).
The smooth function $f(z)$ is approximated by
\begin{equation}\label{eqn:bspln}
\sum_{k=1}^{K_{f}+m_f}b_{K_{f,k}}(z)\gamma_{1k}
=\vb_{K_{f},m_f}(z)\vgamma_1.
\end{equation}
Similarly, functions $g$ and $h$ are approximated using $\vb_{K_{g},m_g}(d)\vbeta_1$ and $\vb_{K_{h},m_h}(d)\valpha_1$.
The smoothness of the functions is controlled through the prior distributions on the basis coefficients by penalising the differences between adjacent coefficients (see Section~\ref{sec:prior}).

Finally, we are interested in estimating the structural effect on the conditional quantile:
\begin{equation}\label{eqn:sq}
S_{y|d}(p)=\beta_0+g(d)+Q_{e|d}(p),
\end{equation}
where $Q_{e|d}(p)$ is the quantile function of $e$.
In addition, since the quantiles of the error term are directly obtained from the nonparametrically modelled distribution, crossing of estimated quantiles is not an issue.
As mentioned above, contrary to the Bayesian nonparametric instrumental variable regression model, since the conditional variance of the error term smoothly varies across the covariate values, the proposed approach models the conditional quantiles of the response variable in a more flexible manner.
Therefore, the smooth effects in the location-scale model, \eqref{eqn:sq}, allows for a great deal of flexibility in the estimated quantile curve.

The present model is an extension of the control function approach of Wiesenfarth~\etal~(2014) to the conditional mean.
This is different from the Bayesian quantile regression approach of Kobayashi~(2016) and Ogasawara and Kobayashi~(2015), who employed Lee's~(2007) control function approach in the quantile regression framework.
The control variable is introduced to the second-stage regression such that the $p$-th quantile for some $p$ of interest is corrected to be equal to zero.
Then, assuming a linear model, the structural effect is estimated from $\tilde{S}_{y,d}(p)=\beta_0+d\delta$.
While this approach works well in several situations, the required assumption for the first-stage error that the $\alpha$-th quantile is equal to zero for some $\alpha\in(0,1)$ can be quite restrictive in modelling the error distribution.
In terms of the quantile modelling, the present approach may be considered a Bayesian nonparametric extension of Gilchrist~(2008), who considered various ways to introduce the covariate influence on the quantile function using parametric models.

\subsection{Default Prior Distributions}\label{sec:prior}
This section discusses the choice of the prior distributions and associated hyperparameters.
For the spline coefficients for the nonlinear effects---$\vgamma_1$, $\vbeta_1$, and $\valpha_1$---the Gaussian random walk priors are assigned following Lang and Brezger~(2004).
In the case of $\vgamma$, the joint prior density is given by
\begin{equation*}
\vgamma_1|\phi_f\propto\exp\left\{-\frac{1}{2\phi_f}\vgamma_1'\mDelta_f \vgamma_1\right\},
\end{equation*}
where $\mDelta_{f}=\mD_{f}'\mD_{f}$ with the difference matrix $\mD_{f}$ and $\phi_{f}$ is the variance parameter that controls the smoothness of the fitted function.
The second-order priors are used throughout this study.
For $\vbeta_1$ and $\valpha_1$, the prior distributions have the parameters $\phi_{g}$ and $\phi_{h}$.
To let the data determine the amount of smoothness, following Wiesenfarth~\etal~(2014), $\phi_f$, $\phi_g$, and $\phi_h$ are estimated along with the coefficients by introducing the vague inverse gamma priors, $\IG(a_\phi, b_\phi)$, where $a_\phi=b_\phi=0.001$.
For the intercepts $\gamma_0$, $\beta_0$, and $\alpha_0$, we adopt the normal priors $\N(0, V_{\vgamma_0})$, $\N(0,V_{\beta_0})$, and $\N(0,V_{\alpha_0})$, respectively.
Since we do not have information on these parameters, the hyperparameters are set such that the priors are vague: $V_{\gamma_0}=V_{\beta_0}=V_{\alpha_0}=100$.

We turn to the prior specification for the components in the Dirichlet process.
Specifically, $\vartheta_1\sim\N(0,T_{10})$ and $\vartheta_2\sim\N(0,T_{20})$, such that the prior means of $\mu_1$ and $\mu_2$ are zero.
For these priors to be vague, we set $T_{10}=T_{20}=100$.
The values of $M_{10}$ and $M_{20}$ are selected such that the prior distributions cover the possible regions where $\mu_1$ and $\mu_2$ are likely to be produced.
Therefore, following Ishwaran and James~(2002), we let $\sqrt{M_{10}}$ and $\sqrt{M_{20}}$ equal four times the sample standard deviations of $d$ and $y$.
The prior specification for the hyperparameters $\vartheta_e$ and $\varphi_e$ of $\eta$ is important because $\eta$ controls the degree of correction required because of the endogeneity.
Since $\eta=\sigma_{12}/\sigma_{11}$, it is natural to assume that $\vartheta_e$ is centred around this value.
Therefore, $\vartheta_e\sim\N(e_0,T_{e0})$ is assumed, where $e_0$ is equal to the sample covariance between $d$ and $y$ divided by the sample variance of $d$ and $T_{e0}=10$.
Since we do not know the variability of $\eta$, that is, the degree of heterogeneity in the endogeneity, we place a prior distribution for $\varphi_e$.
Namely, $\varphi_e\sim\IG(e_1,e_2)$ with $e_1=e_2=2$ such that $\varphi_e$ takes a value between $0$ and $3$ with high probability.
For $\tau$ and $\sigma$, assuming that the data are appropriately rescaled, we set $t_1=t_2=2$, $s_1=2$, and $s_2=1$.
Finally, the precision parameter of the Dirichlet process follows the gamma prior $a\sim\G(a_1,a_2)$ with $a_1=a_2=2$, such that both small and large numbers of mixture components are allowed.
In the simulation study in Section~\ref{sec:sim}, we also consider some alternative prior specifications for $\vartheta_e$ and $a$.

\section{Posterior Inference}\label{sec:mcmc}
\subsection{Markov Chain Monte Carlo}
Posterior inference for the proposed model is based on the output from the MCMC method.
We develop the Gibbs sampling method comprising the slice (Walker,~2007) and retrospective (Papaspiliopoulos and Roberts,~2008) sampler for the variables included in the Dirichlet process and mixture sampler (Chib and Greenberg,~2013) for coefficients in the variance function.

We denote the variable representing the component assigned for the $i$-th observation by $k_i$ for $i=1,\dots,n$ and the weight of the $k$-th component of the mixture by $\pi_k=\Pr(k_i=k)=\omega_k\prod_{l<k}(1-\omega_l)$ with $\omega_k\sim\B(1,a)$, and $\B(a,b)$ is the beta distribution with parameters $a$ and $b$ (Sethuraman,~1994).
We also let $k^*$ denote the minimum integer such that $\sum_{k=1}^{k^*}>1-\min\left\{u_1,\dots,u_n\right\}$.
To apply the slice sampler and retrospective sampler to the variables involved in the Dirichlet process, we work on the following joint density:
\begin{eqnarray}
f(e_i,v_i,u_i)
&=&\sum_{k=1}^\infty I(u_i<\omega_k)
\N(v_i;\mu_{1k},\tau_k) \N(u_i;\mu_{2k}+\eta_k(v_i-\mu_{1k}), s_i\sigma_{k}) \nonumber\\
&=&\sum_{k=1}^\infty I(u_i<\omega_k) \N(\veps_i;\vmu_k,\mSigma_{ki}),
\quad i=1,\dots,n,\nonumber
\end{eqnarray}
where $\N(\cdot;\mu,\sigma)$ denotes the density function of the normal distribution with mean $\mu$ and variance $\sigma$,
$u_i\sim\U(0,1)$, and
$$
\mSigma_{ki}=\left[
\begin{array}{cc}
1&0\\
\eta_{k}&1
\end{array}
\right]\left[
\begin{array}{cc}
\tau_{k}&0\\
0&s_i\sigma_{k}
\end{array}
\right]
\left[
\begin{array}{cc}
1&\eta_{k}\\
0&1
\end{array}
\right].
$$
In addition, collecting terms of \eqref{eqn:mod} and \eqref{eqn:bspln} would be useful for some steps of the Gibbs sampler:
\begin{eqnarray*}
\vy_i&=&\mX_i\tilde{\vbeta}+\veps_i,\quad
\mX_i=\left[
\begin{array}{cc}
\vz_i'&\zero\\
\zero&\vx_i'
\end{array}
\right], \quad
\tilde{\vbeta}=(\vgamma',\vbeta')',\quad
\veps_i\sim\N(\vmu_{k_i},\mSigma_{k_i i}),
\end{eqnarray*}
where
$\vy_i=(d_i,y_i)'$,
$\vz_i=(1,\vb_{K_f,m_f}(z_i))'$,
$\vgamma=(\gamma_0,\vgamma_1')'$,
$\vx_i=(1,\vb_{K_{g},m_g}(d_i))'$,
$\vbeta=(\vbeta_0,\vbeta_1')'$,
$s_i=\exp(\vw_i'\valpha)$,
$\vw_i=(1, \vb_{K_{h},m_h}(d_i))'$
$\valpha=(\alpha_0,\valpha_1')'$, and
$\vmu_{k_i}=(\mu_{1k_i},\mu_{2k_i})'$.

Our Gibbs sampler proceeds by alternately sampling from the full conditional distributions of $\{u_i\}_{i=1}^n$, $\{\omega_l\}_{l=1}^{k^*}$, $\{k_i\}_{i=1}^n$, $\{\vmu_k\}_{k=1}^{k^*}$, $\{\eta_{k}\}_{k=1}^{k^*}$, $\{\tau_k\}_{k=1}^{k^*}$, $\{\sigma_k\}_{k=1}^{k^*}$, $a$, $\vartheta_1$, $\vartheta_2$, $\vartheta_e$, $\varphi_e$, $\tilde{\vbeta}$, $\valpha$, $\phi_f$, $\phi_g$, and $\phi_h$.

\begin{itemize}
\item\textbf{Sampling $\{u_i\}_{i=1}^n$:} Generate $u_i$ from $\U(0,\pi_{k_i})$ for $i=1,\dots,n$.

\item\textbf{Sampling $\{\omega_l\}_{k=1}^{k^*}$:} Generate $\omega_k$ from $\B(1+n_k,n-\sum_{l\leq k}n_l+a)$, where $n_k=\sum_{i=1}^n I(k_i=k)$ for $k=1,\dots,k^*$.
\item\textbf{Sampling $\{k_i\}_{i=1}^n$:} Generate $k_i$ from the multinomial distribution with probabilities
\begin{equation*}
\Pr(k_i=k)\propto \N(\vy_i;\vmu_k+\mX_i\tilde{\vbeta},\mSigma_{ki}) I(u_i<\pi_k),\quad k=1,\dots,k^*.
\end{equation*}

\item{\textbf{Sampling $a$:}} By introducing $r\sim\B(a+1,n)$, the full conditional distribution of $a$ is a mixture of two gamma distributions given by
\begin{equation*}
\varphi\G(a_1+n^*,a_2-\log r)+(1-\varphi)\G(a_1+n^* -1, a_2-\log r),
\end{equation*}
where $n^*$ is the number of distinct clusters and $\varphi/(1-\varphi)=(a_1+n^*-1)/(n(a_2-\log r))$.
See Escober and West~(1995).

\item\textbf{Sampling $\{\vmu_k\}_{k=1}^{k^*}$:}
For $k=1,\dots,k^*$, $(\mu_{1k},\mu_{2k})'$ is sampled from $\N(\hat{\vmu}_k,\hat{\mM}_k)$, where
\begin{equation*}
\hat{\mM}_k=\left[\sum_{i:k_i=k}\mSigma_{ki}^{-1}+\mM_0^{-1}\right]^{-1}, \quad
\hat{\vmu}_k=\hat{\mM}_k\left[\sum_{i:k_i=k} \mSigma_{ki}^{-1}(\vy_i-\mX_i\tilde{\vbeta})+\mM_0^{-1}\mathbf{\vartheta} \right],
\end{equation*}
where $\mM_0=\diag(M_1,M_2)$ and $\mathbf{\vartheta}=(\vartheta_1,\vartheta_2)'$.

\item\textbf{Sampling $\{\eta_{k}\}_{k=1}^{k^*}$:}
For $k=1,\dots,k^*$, $\eta_k$ is sampled from $\N(\hat{\eta}_k, \hat{E}_k)$, where
\begin{equation*}
\hat{E}_k=\left[\sum_{i:k_i=k}\frac{(v_i-\mu_{1k})^2}{\sigma_k\exp(\vw_i'\valpha)}+\frac{1}{\varphi_e}\right]^{-1}, \quad
\hat{\theta}_k=\hat{E}_k\left[\sum_{i:k_i=k}\frac{(v_i-\mu_{1k})(y_i-\mu_{2k}-\vx_i'\vbeta)}{\sigma_k\exp(\vw_i'\valpha)}+\frac{\vartheta_e}{\varphi_e}\right].
\end{equation*}

\item\textbf{Sampling $\{\tau_{k}\}_{k=1}^{k^*}$:}
For $k=1,\dots,k^*$, $\tau_k$ is sampled from $\IG(\hat{t}_{1k}, \hat{t}_{2k})$, where
\begin{equation*}
\hat{t}_{1k}=t_1+\frac{n_k}{2},\quad
\hat{t}_{2k}=t_2+\frac{1}{2}\sum_{i:k_i=k}(d_i-\mu_{1k}-\vz_i'\vgamma)^2.
\end{equation*}

\item\textbf{Sampling $\{\sigma_{k}\}_{k=1}^{k^*}$:}
For $k=1,\dots,k^*$, $\sigma_k$ is sampled from $\IG(\hat{s}_{1k}, \hat{s}_{2k})$, where
\begin{equation*}
\hat{s}_{1k}=s_1+\frac{n_k}{2},\quad
\hat{s}_{2k}=s_2+\frac{1}{2}\sum_{i:k_i=k}\frac{(y_i-\mu_{2k}-\vx_i'\vbeta-\eta_k(v_i-\mu_{2k}))^2}{\exp(\vw_i'\valpha)}.
\end{equation*}

\item\textbf{Sampling $\vartheta_1$ and $\vartheta_2$:}
The full conditional distribution of $\vartheta_1$ is $\N(\hat{\vartheta}_1,\hat{T}_1)$, where
\begin{equation*}
\hat{T}_1=\left(\frac{k^*}{M_{10}}+\frac{1}{T_{10}}\right)^{-1},\quad
\hat{\vartheta}_1=\frac{\hat{T}_1}{M_{10}}\sum_{k=1}^{k^*}\mu_{1k}.
\end{equation*}
The sampling for $\vartheta_2$ is performed in a similar manner.

\item\textbf{Sampling $\vartheta_e$ and $\varphi_e$:}
The full conditional distribution of $\vartheta_e$ is $\N(\hat{\vartheta}_e,\hat{T}_e)$, where
\begin{equation*}
\hat{T}_e=\left(\frac{k^*}{\varphi_e} + \frac{1}{T_{e0}}\right)^{-1},\quad
\hat{\vartheta}_e=\hat{T}_e\left(\sum_{k=1}^{k^*}\frac{\eta_k}{\varphi_e}+\frac{e_0}{T_{e0}}\right),
\end{equation*}
and the full conditional distribution of $\varphi_e$ is $\IG(\hat{e}_1, \hat{e}_2)$, where
\begin{equation*}
\hat{e}_1=e_1+\frac{k^*}{2},\quad
\hat{e}_2=e_2+\sum_{k=1}^{k^*}\frac{(\eta_k-\vartheta_e)^2}{2}.
\end{equation*}

\item\textbf{Sampling $\tilde{\vbeta}$:}
The full conditional distribution of $\tilde{\vbeta}$ is denoted by $\N(\hat{\tilde{\vbeta}},\hat{\tilde{\mB}})$, where
\begin{equation*}
\hat{\tilde{\mB}}=\left[\sum_{i=1}^n\mX_i'\mSigma_{k_i i}^{-1}\mX_i+\tilde{\mB}_0^{-1}\right]^{-1}, \quad
\hat{\tilde{\vbeta}}=\hat{\tilde{\mB}}\left[\sum_{i=1}^n\mX_i'\mSigma_{k_i i}^{-1}(\vy_i-\vmu_{k_i})+\tilde{\mB}_0^{-1}\tilde{\vb}_0\right],
\end{equation*}
$\tilde{\vb}_0=(0,\zero',0,\zero')'$, and $\tilde{\mB}_0^{-1}$ is the block diagonal matrix with $V^{-1}_{\gamma_0}$, $\phi_f^{-1}\mDelta_{f}$, $V^{-1}_{\beta_0}$, and $\phi_g^{-1}\mDelta_{g}$ on the diagonal blocks.

\item\textbf{Sampling $\valpha$:}
Note that
\begin{equation*}
y_i^*=\log(y_i-\mu_{2k_i}-\vx_i'\vbeta-\eta_{k_i}(v_i-\mu_{1k_i}))^2 - \log(\sigma_{k_i})
=\vw_i'\valpha+e_i^*,\quad i=1,\dots,n,
\end{equation*}
where $e_i^*=\log e_i^2$ follows the log-$\chi^2$ distribution with one degree of freedom.
It is known that the log-$\chi^2$ distribution can be accurately approximated by the ten-component mixture of normal distributions
\begin{equation*}
f(e^*)\approx\sum_{h=1}^{10}\varrho_h\N(e^*;\xi_h,\zeta_h),
\end{equation*}
 with the known means $\xi_h$, variances $\zeta_h$, and component weights $\varrho_h$, $h=1,\dots,10$ (Omori~\etal,~2007).
The values for $\xi_h$, $\zeta_h$, and $\varrho_h$ are summarised in \ref{sec:appa}.
Introducing the indicator $h_i\in\{1,\dots,10\}$, the mixture sampler proceeds by first sampling $h_i$ for $i=1,\dots,n$ from the multinomial distribution with probabilities
\begin{equation*}
\Pr(h_i=h)\propto \varrho_h\N(y_i^*;\xi_h+\vw_i'\alpha,\zeta_h), \quad h=1,\dots,10,
\end{equation*}
and then, sampling $\valpha$ from $\N(\hat{\valpha},\hat{\mA})$, where
\begin{equation*}
\hat{\mA}=\left[\sum_{i=1}^n\frac{\vw_i\vw_i'}{\zeta_{h_i}}+\tilde{\mA}_0^{-1}\right]^{-1},\quad
\hat{\valpha}=\hat{\mA}\left[\sum_{i=1}^n\frac{\vw_i(y_i^*-\xi_{h_i})}{\zeta_{h_i}}+\tilde{\mA}_0^{-1}\tilde{\va}_0\right],
\end{equation*}
$\tilde{\va}_0=(0,\zero')'$, and $\tilde{\mA}_0^{-1}$ is the block diagonal matrix with $V^{-1}_{\alpha_0}$ and $\phi_h^{-1}\mDelta_{h}$ on the diagonal blocks.
The values for $\xi_h$, $\zeta_h$, and $\varrho_h$ are presented in \ref{sec:appa}.
See also Chib and Greenberg~(2013).

\item\textbf{Sampling $\phi_f$, $\phi_g$, and $\phi_h$:}
The full conditional distribution of $\phi_f$ is $\IG(\hat{a}_{\phi_f},\hat{b}_{\phi_f})$, where
\begin{equation*}
\hat{a}_{\phi_f}=a_\phi+0.5 \text{rank}(\mDelta_f),\quad
\hat{b}_{\phi_f}=b_\phi+0.5\vgamma_1'\mDelta_{f}\vgamma_1.
\end{equation*}
We sample $\phi_g$ and $\phi_h$ in the same manner.

\end{itemize}

\subsection{Quantile Estimation}
The $t$-th iteration of the Gibbs sampler provides a posterior draw from $G$:
\begin{equation*}
G^{(t)}=\sum_{k=1}^{k^*} \pi_k^{(t)} \delta_{\vtheta_k^{(t)}}+\left(1-\sum_{k=1}^{k^*}\pi_k^{(t)}\right)G_0,
\end{equation*}
where superscript $(t)$ denotes the sampled value at the $t$-th iteration and $\delta_{\va}$ is the Dirac mass at $\va$.
The density estimate for the error distribution can be obtained from the posterior predictive density.
Then, given a draw $G^{(t)}$ from the posterior distribution, the posterior predictive density of the error term is computed as
\begin{equation*}
f^{(t)}(\veps|\vy)=\sum_{k=1}^{k^*}\pi_k^{(t)}f(\veps|\vtheta_k^{(t)})+\left(1-\sum_{k=1}^{k^*}\pi_k^{(t)}\right)\int f(\veps|\vtheta)G_0(\d\vtheta).\end{equation*}
The second term is approximated by
\begin{equation*}
\left(1-\sum_{k=1}^{k^*}\pi_k^{(t)}\right)\frac{1}{M}\sum_{m=1}^M f(\veps|\vtheta^{(m)}),
\end{equation*}
where $\vtheta^{(m)}$ is the $m$-th draw from $G_0$ and $M$ can be as large as $10$ (see, \eg\ Jin and Maheu,~2016).
Over the $T$ iterations of the Gibbs sampler, the error density is estimated using
\begin{equation*}
f(\veps|\vy)\approx\frac{1}{T} \sum_{t=1}^T f^{(t)}(\veps|\vy).
\end{equation*}

To estimate \eqref{eqn:sq} for any $p\in(0,1)$, we obtain draws $\vx'\vbeta^{(t)}+Q^{(t)}_{e|d}(p)$ from the posterior distribution,
where $Q^{(t)}_{e|d}(p)$ is a posterior draw of the $p$-th quantile of $e$ and computed from the posterior draw of the distribution function denoted by
\begin{equation*}
F^{(t)}(e|d)\approx\sum_{k=1}^{k^*}\pi_{k}^{(t)}\Phi\left(\frac{e-\mu_{2k}^{(t)}}{\sqrt{s\sigma_k^{(t)}}}\right)+
\left(1-\sum_{k=1}^{k^*}\pi_k^{(t)}\right)\frac{1}{M}\sum_{m=1}^M\Phi\left(\frac{e-\mu_{2k}^{(m)}}{\sqrt{s\sigma_k^{(m)}}}\right),
\end{equation*}
where $\Phi(\cdot)$ denotes the distribution function of the standard normal distribution.
As in Taddy and Kottas~(2010), $F(e|d)$ and $Q_{e|d}(p)$ must be computed over a sufficiently fine grid of covariate values.
For a high dimensional covariate vector, the inference would be feasible only when a one- or two-dimensional grid is used for the covariates of interest, while the other variables are fixed at some value.

\section{Simulation Study}\label{sec:sim}
\subsection{Simple Model}\label{sec:sim1}
To illustrate the importance of modelling the variance function for quantile inference, the data are generated using a simple linear model:
\begin{equation*}
\begin{split}
d_i&=\gamma_0+ \gamma_1z_i + v_i,\quad v_i\sim\N(\mu_{1i},\tau_i)\\
y_i&=\beta_0+\beta_1 d_i + \eta_i (v_i-\mu_1) + e_i,\quad e_i\sim\N(\mu_{2i},s_i\sigma_i), \\
s_i&=\exp\{\alpha_0+\alpha_1d_i\},
\end{split}
\end{equation*}
for $i=1,\dots,n$, where $n=200$ and $1000$, $\gamma_0=0$, $\gamma_1=2$, $\beta_0=1$, $\beta_1=1$, $\alpha_0=0$, and $\alpha_1=0.3$, and $(\mu_{1i},\mu_{2i},\eta_i,\tau_i,\sigma_i)$ takes the value of $(1,-1,2.0,1,1)$, $(0,0,0.3,1,1)$, or $(-1,1,0.8,1,1)$ with equal probabilities, such that the error distribution is a three-component mixture of normal distributions.
This specification for the mixture components implies that the correlation coefficients between $v$ and $e$ are $0.894$, $0.287$, and $0.624$.
The data are replicated 100 times.

We fit a simple version of the proposed model comprising a linear mean and variance functions but with a nonparametric error distribution.
For comparison, we also fit two alternative models.
The first is denoted by the restricted model and set to $s_i=1,\ i=1,\dots,n,$, such that the conditional variance is no longer a function of the covariates, following Conley~\etal's~(2008) model and Wiesenfarth~\etal's~(2014) simplified version.
The second model is denoted by the uncorrected model and ignores the endogeneity of $d$ but accounts for covariate dependent variance.
Note that due to the bimodality of the error density, Kobayashi's~(2016) approach is not applicable because it assumes a unimodal density for the first-stage error.
For the three models, 10,000 draws are obtained from the posterior distribution using a Gibbs sampler after discarding the initial 1,000 draws as a burn-in period.
Every fifth draw is retained for posterior estimation.

\begin{figure}[H]
\centering
\includegraphics[scale=0.2]{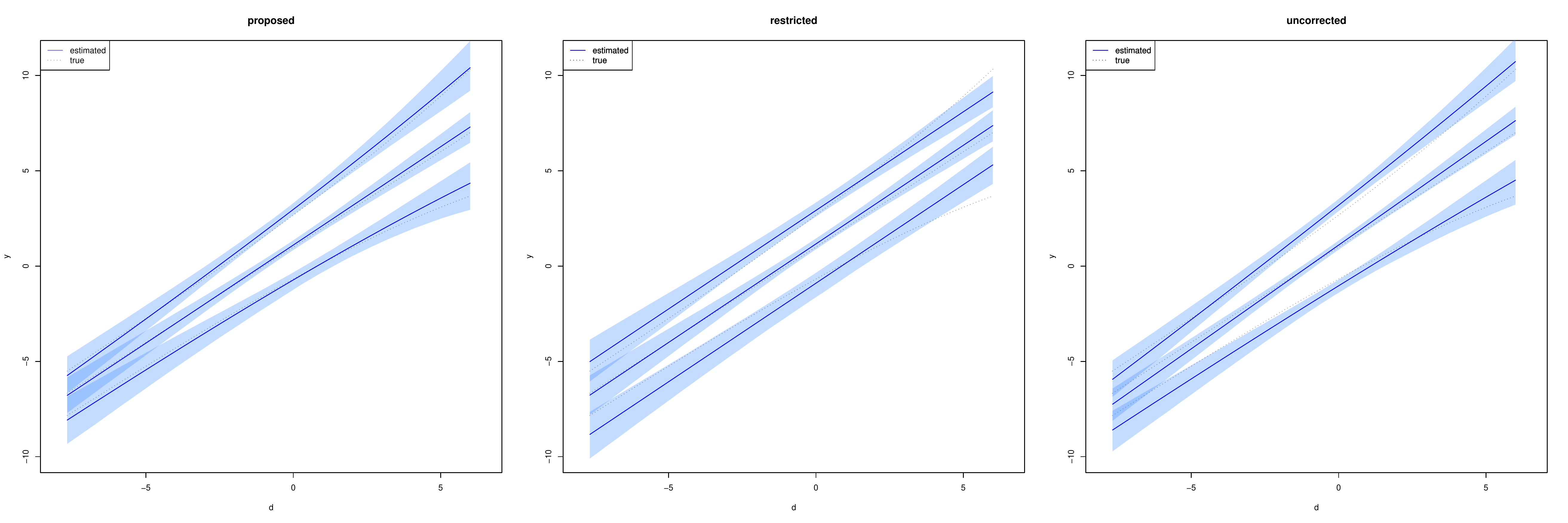}\\
\includegraphics[scale=0.2]{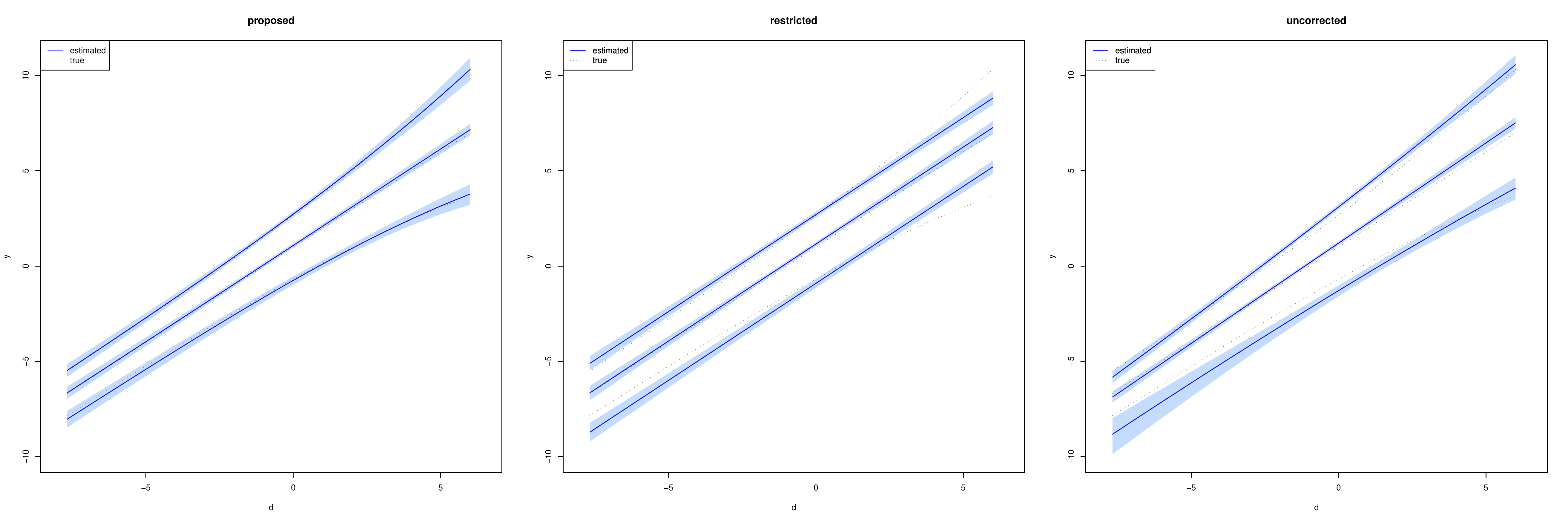}
\caption{Simple model: posterior means and 95\% credible intervals for $S_{y|d}(p)$ for $p=0.1$, $0.5$, and $0.9$ under the proposed, restricted, and uncorrected models for $n=200$ (top row) and $n=1000$ (bottom row)}
\label{fig:sim1a}
\end{figure}

Figure~\ref{fig:sim1a} presents the typical result for the posterior means and 95\% credible intervals for the $p$-th quantile curves \eqref{eqn:sq} with $p=0.1$, $0.5$, and $0.9$ for the proposed, restricted, and uncorrected models for $n=200$ and $1000$.
In the case of $n=200$, the figure shows that the true curves are not included in the 95\% credible intervals for the restricted and uncorrected models for some parts of the region where $d$ is observed and for some $p$.
Increasing the sample size to $n=1000$ shows a clear difference between the three models.
The quantile curves obtained from the restricted model must be parallel to each other and the 95\% credible intervals do not include the true curves for most parts of the region where $d$ is observed, especially for $p=0.1$ and $0.9$.
For the uncorrected model, the figure shows that ignoring endogeneity leads to a bias and the 95\% credible intervals do not include the true curves for most of the region for all three quantile levels considered here.
In contrast, the proposed model correctly estimates the true quantile curves over the entire region where $d$ is observed.

\begin{figure}[H]
\centering
\includegraphics[scale=0.2]{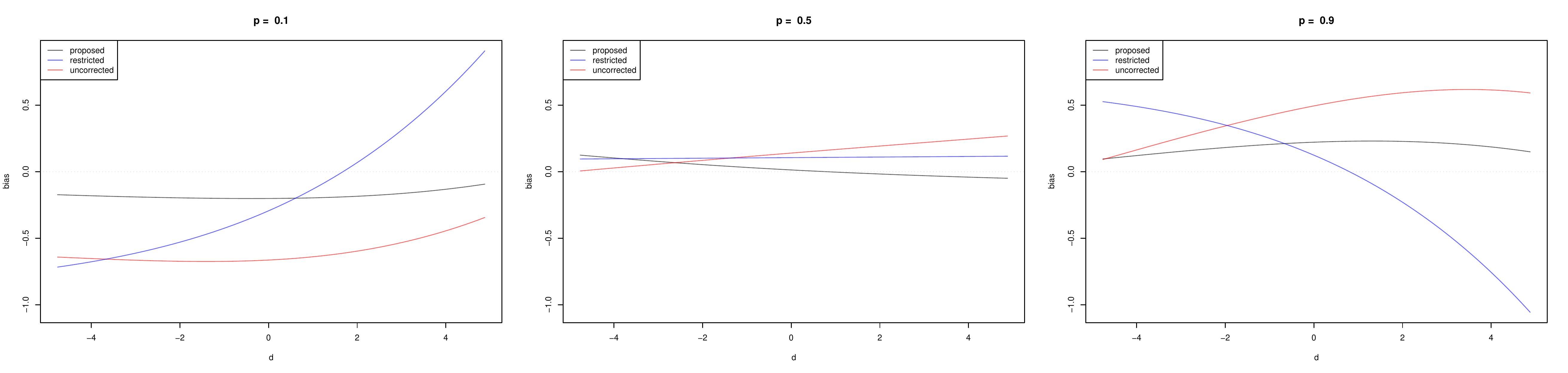}\\
\includegraphics[scale=0.2]{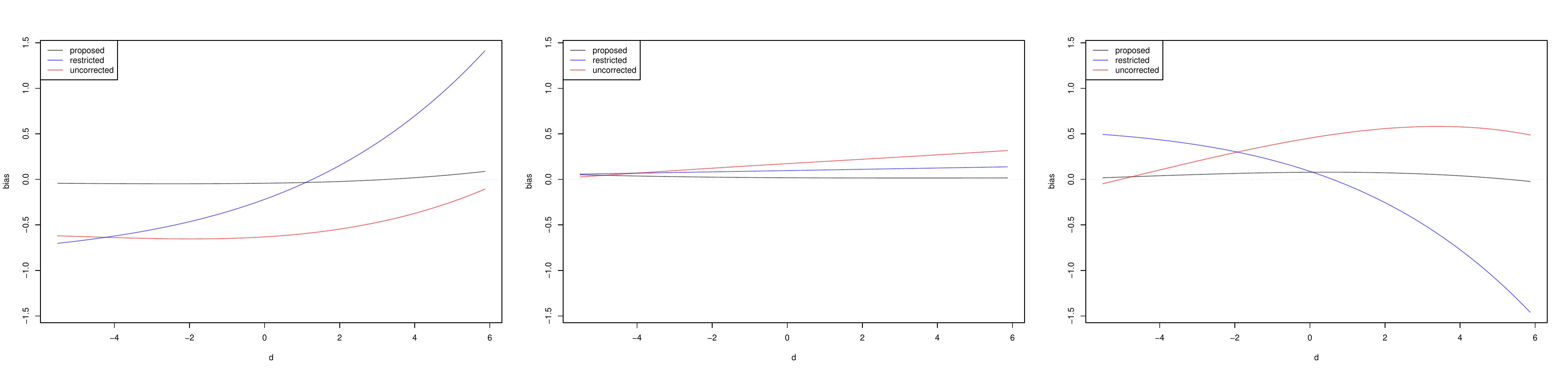}
\caption{Simple model: biases for the proposed, restricted, and uncorrected models for $n=200$ (top row) and $n=1000$ (bottom row)}
\label{fig:sim1b}
\end{figure}

\begin{figure}[H]
\centering
\includegraphics[scale=0.2]{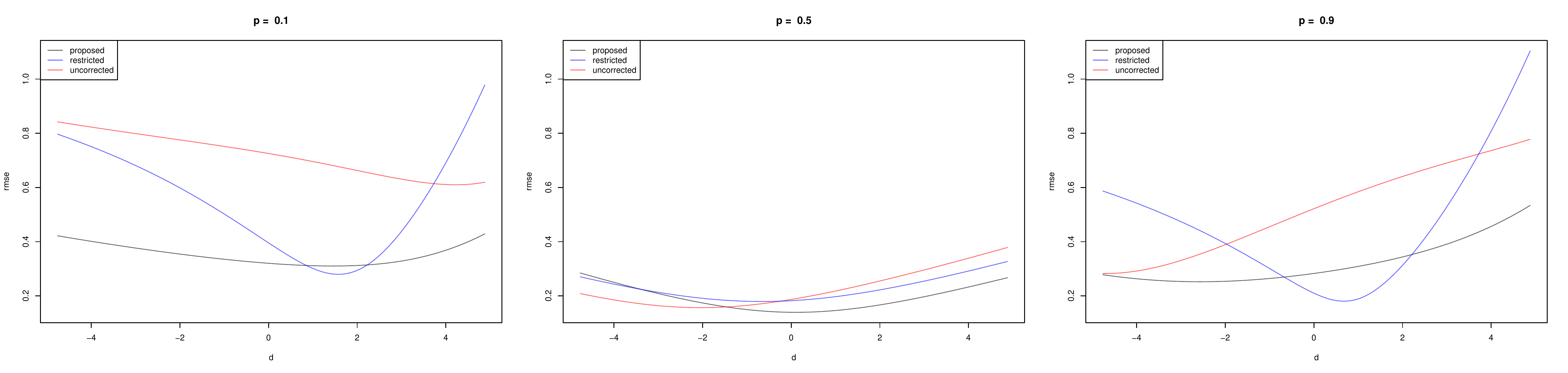}\\
\includegraphics[scale=0.2]{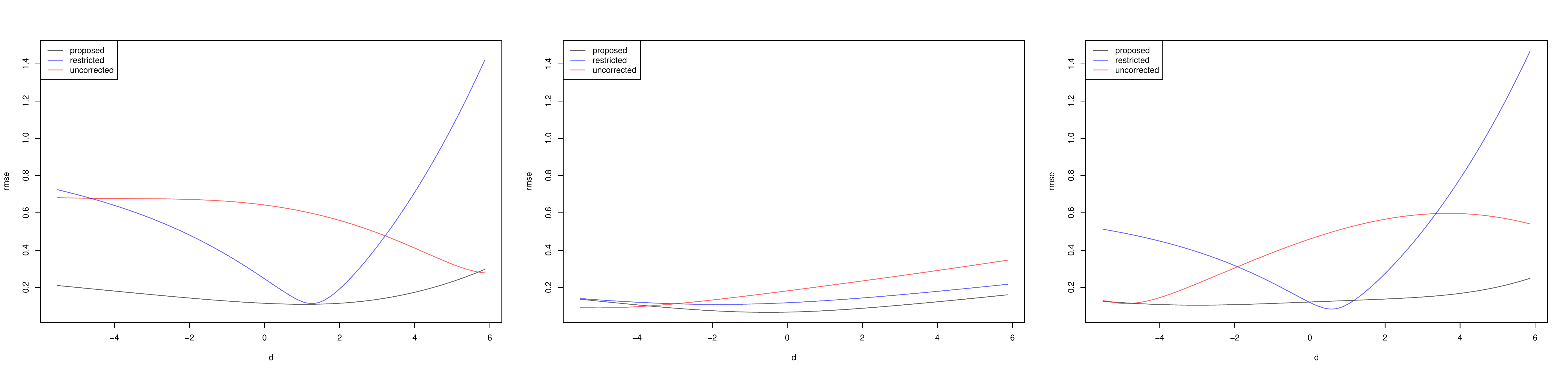}
\caption{Simple model: RMSEs for the proposed, restricted, and uncorrected models for $n=200$ (top row) and $n=1000$ (bottom row)}
\label{fig:sim1c}
\end{figure}

Figures~\ref{fig:sim1b} and \ref{fig:sim1c} present the biases and root mean squared errors (RMSEs) for the proposed, restricted, and uncorrected models for $p=0.1$, $0.5$, and $0.9$, respectively.
The bias and RMSEs are defined as
\begin{equation*}
\text{bias}(d_g)=\frac{1}{R}\sum_{r=1}^R\left(\hat{S}^{(r)}_{y|d_g}(p)-S_{y|d_g}(p)\right),\quad
\text{RMSE}(d_g)=\sqrt{\frac{1}{R}\sum_{r=1}^R\left(\hat{S}^{(r)}_{y|d_g}(p)-S_{y|d_g}(p)\right)^2},
\end{equation*}
over $R=100$ replication and over the grid $d_g, g=1,\dots,G$ on the space of $d$,
where $\hat{S}_{y|d_g}^{(r)}(p)$ is the posterior mean of \eqref{eqn:sq} at $d_g$ for the $r$-th replication.
We set $G=100$ and let $d_1=\max_r\{\min_i\{ d_i^{(r)}\}\}$ and $d_{100}=\min_r\{\max_i\{d_i^{(r)}\}\}$.
Figure~\ref{fig:sim1b} shows that the biases for the restricted and uncorrected models nonlinearly depend on the value of $d$ and do not become smaller as the sample size increases.
By contrast, for all quantiles, the biases for the proposed model are near zero for the region considered and become closer to zero as the sample size increases.
Similar to the biases, Figure~\ref{fig:sim1c} shows the clear dependence of RMSEs for the restricted and uncorrected models on the value of $d$.

\subsection{Nonparametric Models}\label{sec:sim2}
In this section, the proposed model is demonstrated under various settings for the mean and variance functions.
The data are generated from
\begin{equation*}
\begin{split}
d_i&=g(z_i) + v_i, \quad v_i\sim\N(\mu_{2i},\tau_i),\\
y_i&=g(d_i) + \eta_i (v_i-\mu_1) + e_i,\quad e_i\sim\N(\mu_{1i},s(d_i)\sigma_i),
\end{split}
\end{equation*}
where we use the same setting for the error distribution as in Section~\ref{sec:sim1}.
For the mean and variance functions, the following three settings are considered:
\renewcommand{\labelenumi}{(\roman{enumi})}
\begin{enumerate}
\item
$f(z)=1.5z+2\exp(-16z^2)$, \
$g(d)=1+\sin(d)$,  \
$s(d)=\exp(0.5-0.3d^2),$

\item
$g(z)=\exp(-0.5z)$, \
$f(d)=x-0.5x^2$,  \
$s(d)=3\N(d;1,0.5)+\N(d;-1,2)+0.1,$

\item
$f(z)=2z$, \
$g(d)=3\Phi(d/\sqrt{2})$,  \
$s(d)=\exp(0.2d)$
\end{enumerate}
Similar settings can be found in, for example, Chib and Greenberg~(2010; 2013) and Leslie~\etal~(2007).
We consider two cases: $n=500$ and $2000$.
The data are replicated 100 times.

To estimate the mean and variance functions, we use the cubic B-spline functions with equidistant knots by setting
the first and last interior knots equal to the values of minimum and maximum observations.
In the case of $z$, for example,
$\kappa^{(r)}_{f,1}=\min_i\{z_i^{(r)}\}$ and $\kappa^{(r)}_{f,2m_f+K_{f}+1}=\max_i\{z^{(r)}_{i}\}$ for the $r$-th replication.
Following the rule of thumb, we set $K_f=K_g=K_h=\min\{40,n/4\}$ as in Kauermann and Wegener~(2011) and Wiesenfarth~\etal~(2014).

As in the previous simulation, the performance of the proposed, restricted, and uncorrected models are compared.
In addition, the following two alternative prior specifications are applied to the error distribution in the proposed model.
As mentioned in Section~\ref{sec:model}, $\eta$ is an important quantity controlling the degree of correction required from the endogeneity in $d$.
The precision parameter $a$ controls how close the realisation of $G$ is to base measure $G_0$.
Therefore, we consider two ways of introducing prior information with respect to $\eta$ and $a$ and compare the results.
The first alternative prior deflates the prior variance of $\eta$ by about five times by setting $T_{e0}=1$, $e_1=2$, and $e_1=1$ and inflates the prior expectation of $a$ by about five times by setting $a_1=2$ and $a_2=0.4$.
The second alternative prior inflates the prior variance of $\eta$ by about five times by setting $T_{e0}=60$ and $e_1=e_2=2$ and deflates the prior expectation of $a$ by about five times by setting $a_1=2$ and $a_2=10$.

\begin{figure}[H]
\centering
\includegraphics[width=\textwidth]{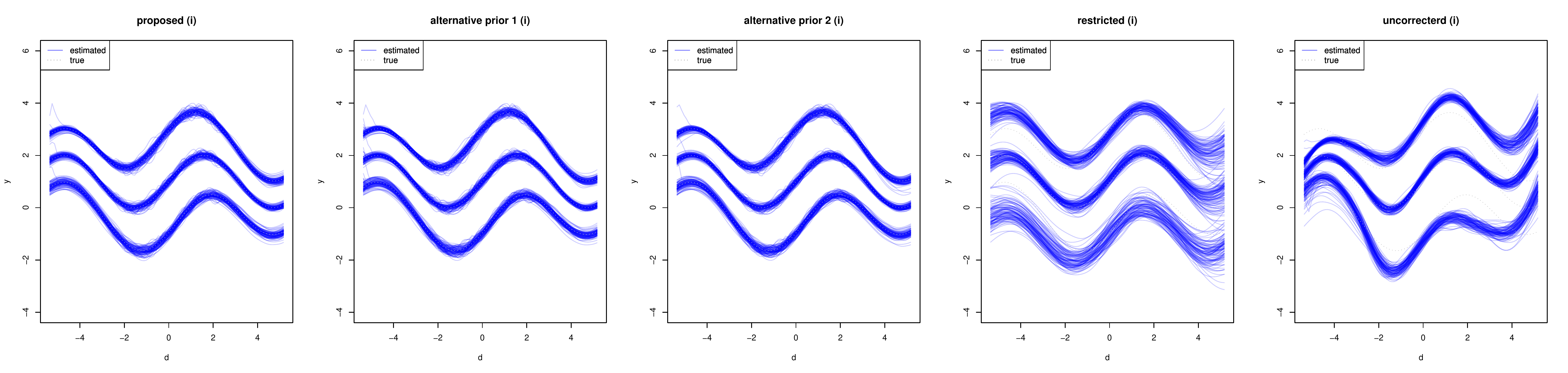}\\
\includegraphics[width=\textwidth]{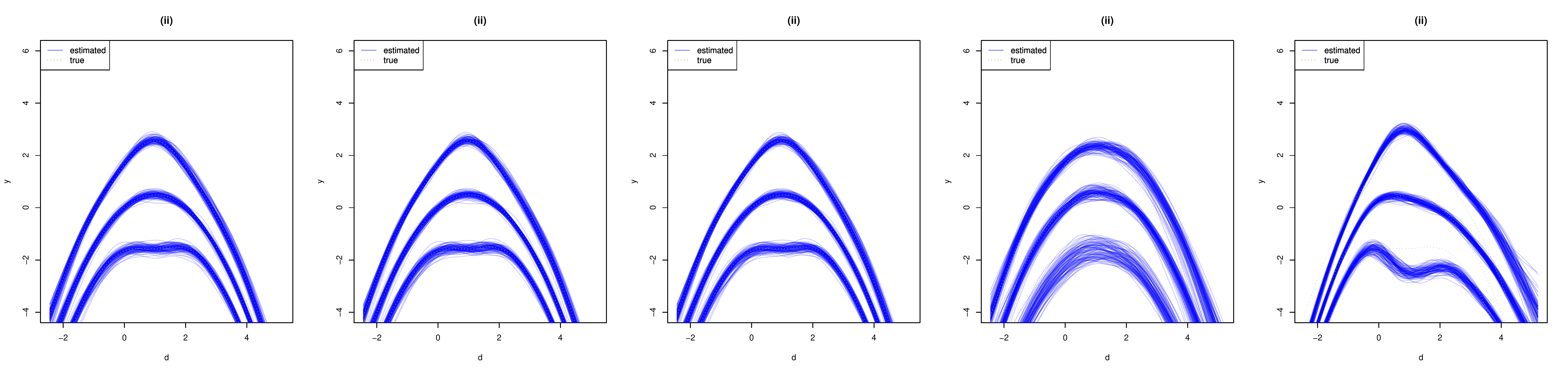}\\
\includegraphics[width=\textwidth]{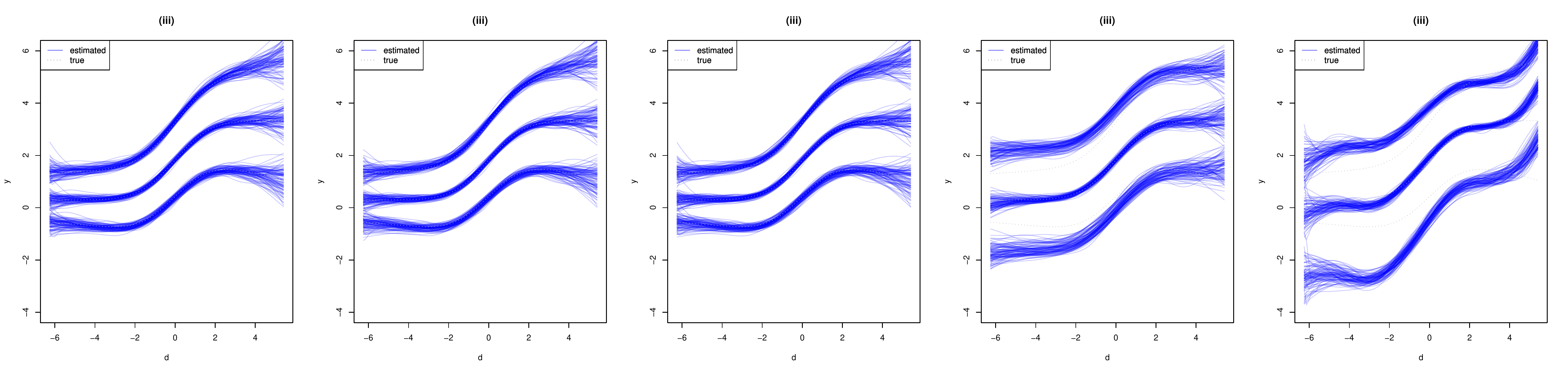}
\caption{Nonparametric models: 100 estimated curves for Settings~(i), (ii), and (iii) for $n=2000$ }
\label{fig:sim2c}
\end{figure}

\begin{figure}[H]
\centering
\includegraphics[scale=0.2]{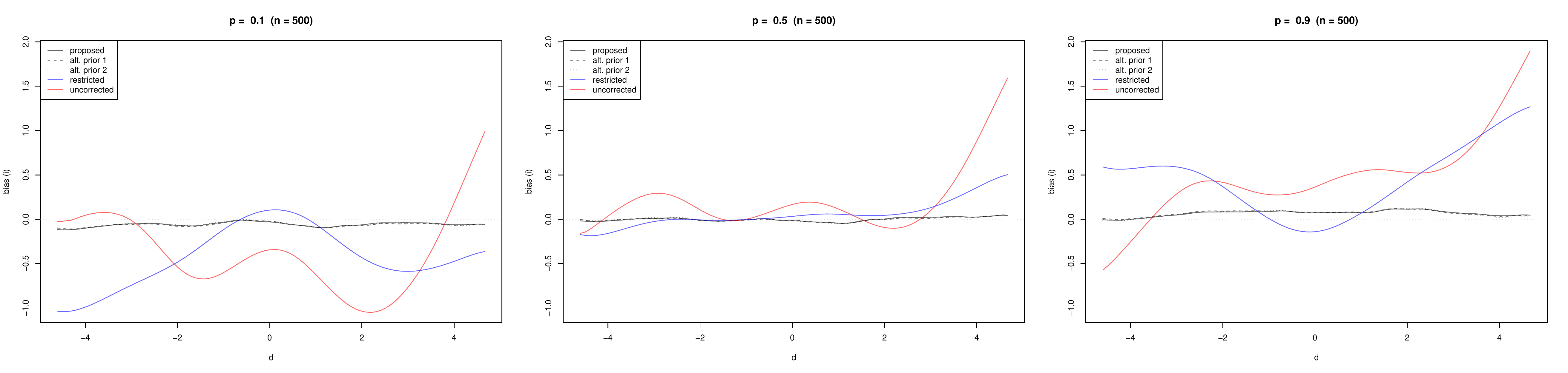}\\
\includegraphics[scale=0.2]{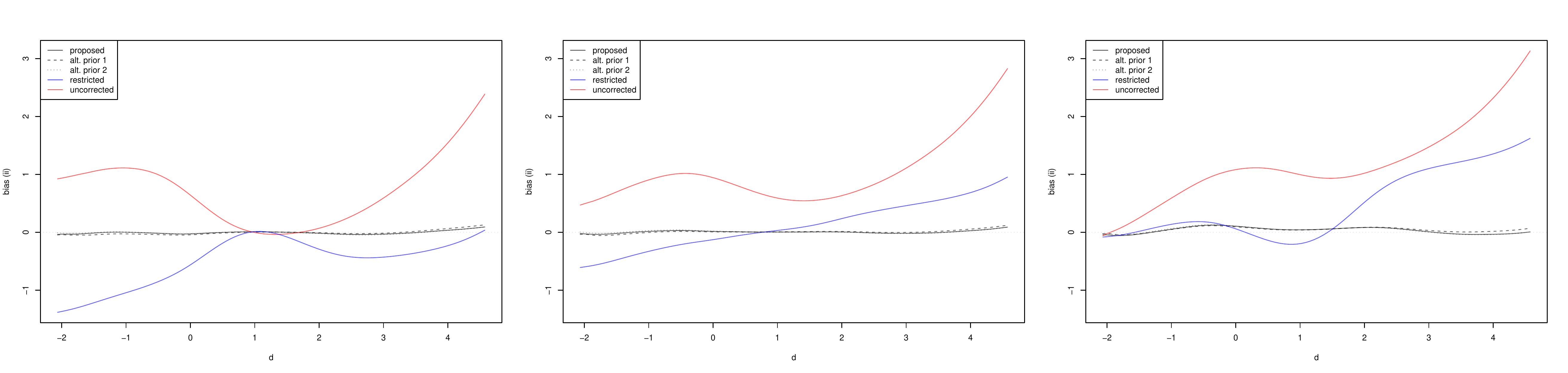}\\
\includegraphics[scale=0.2]{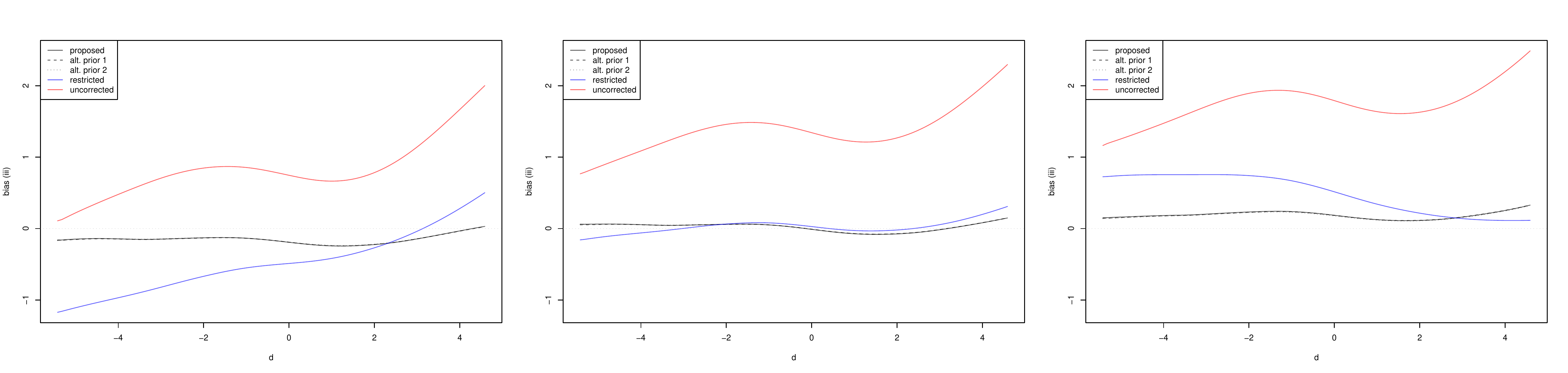}\\
\includegraphics[scale=0.2]{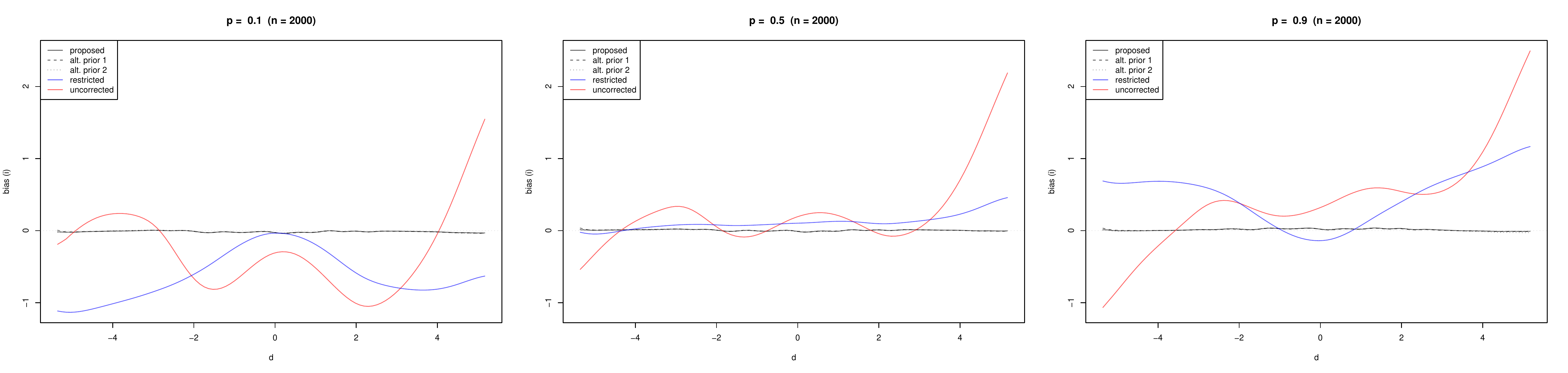}\\
\includegraphics[scale=0.2]{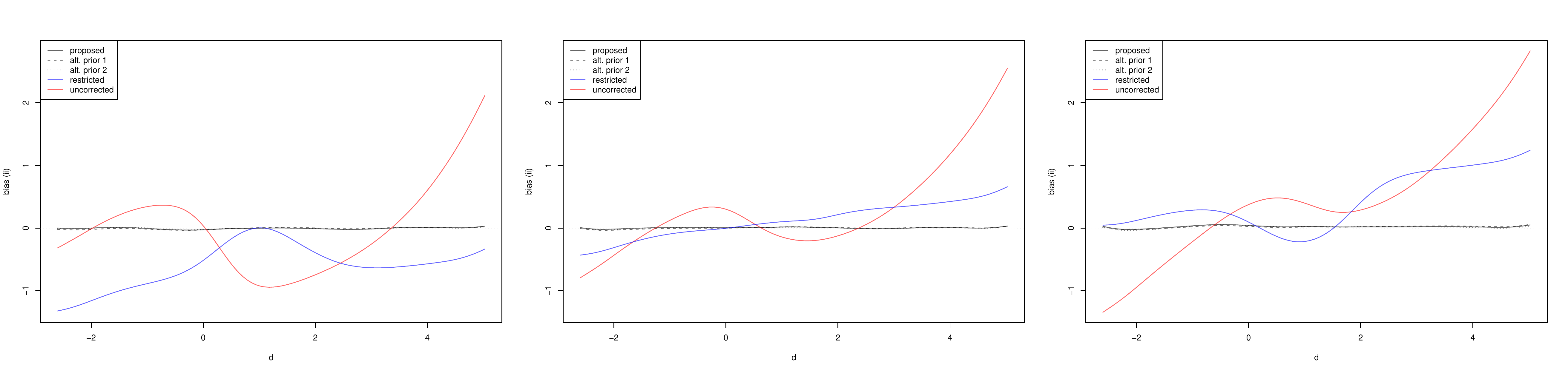}\\
\includegraphics[scale=0.2]{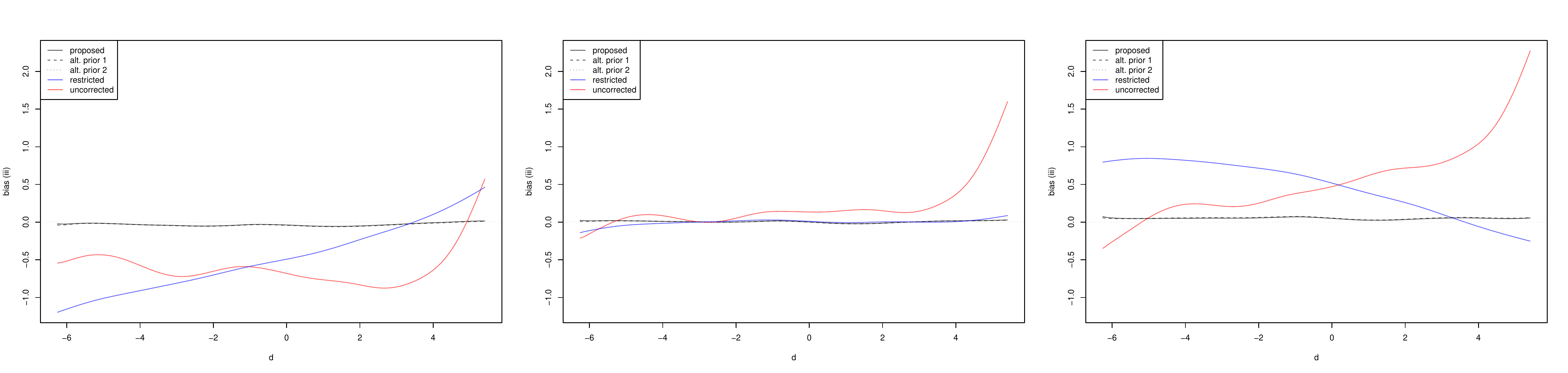}
\caption{Nonparametric models (i), (ii), and (iii): biases for the proposed, restricted, and uncorrected models for $n=500$ (top three rows) and $n=2000$ (bottom three rows)}
\label{fig:sim2a}
\end{figure}

\begin{figure}[H]
\centering
\includegraphics[scale=0.2]{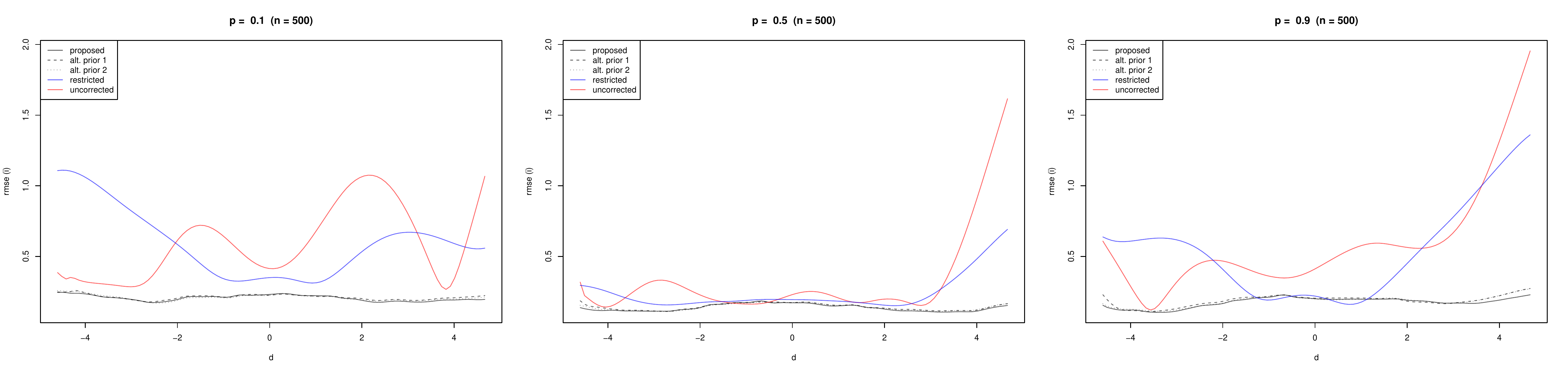}\\
\includegraphics[scale=0.2]{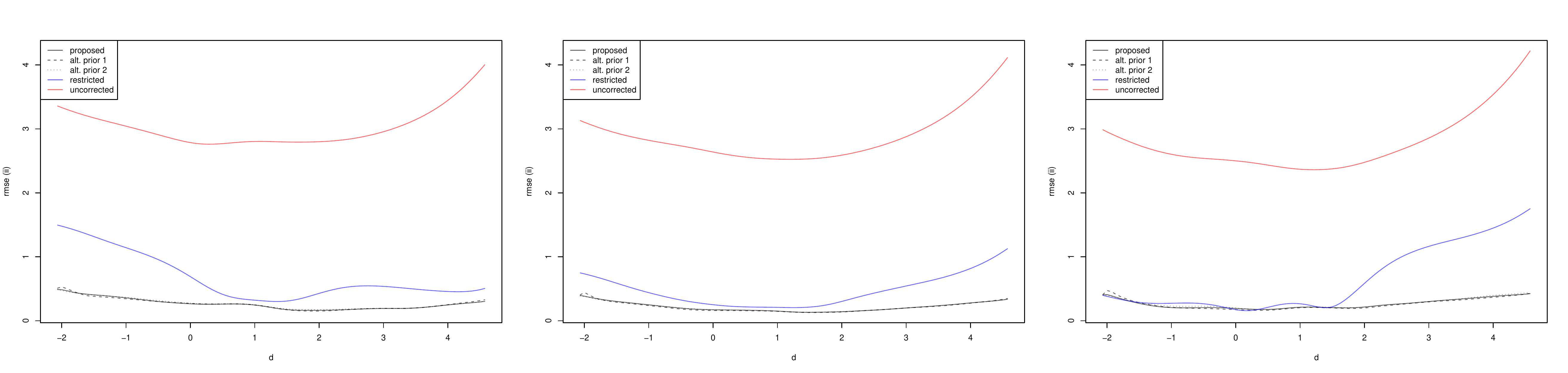}\\
\includegraphics[scale=0.2]{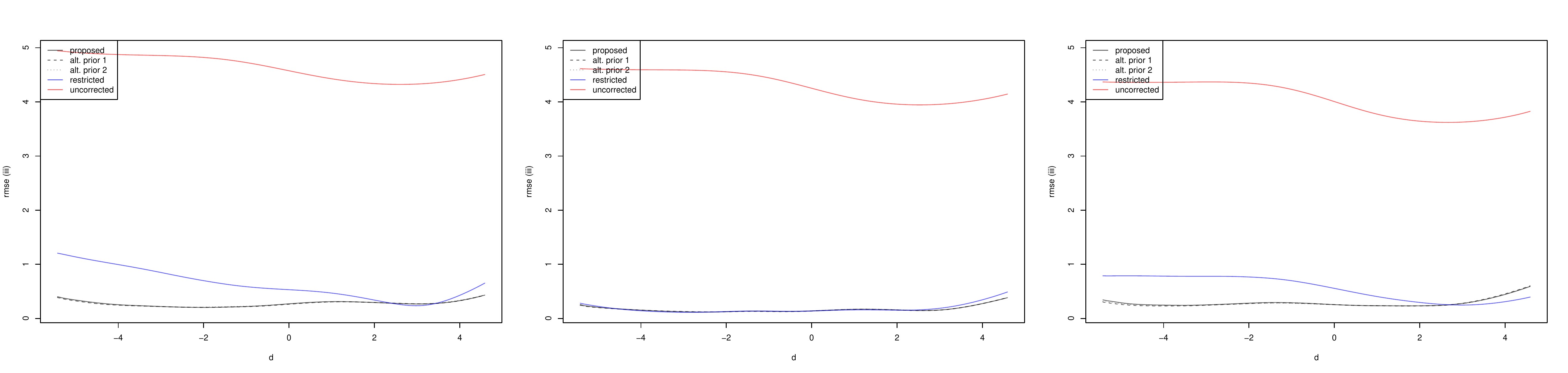}\\
\includegraphics[scale=0.2]{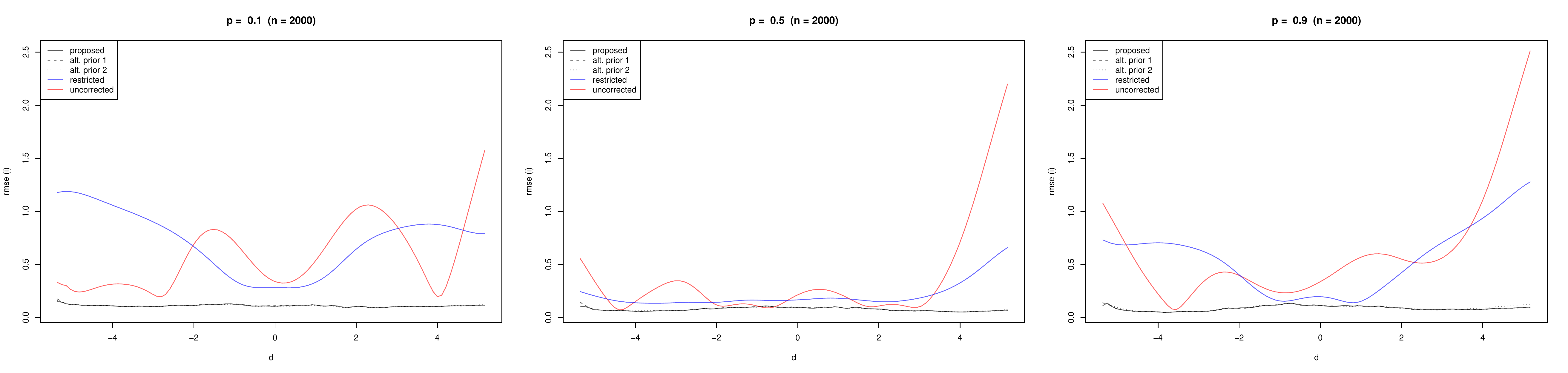}\\
\includegraphics[scale=0.2]{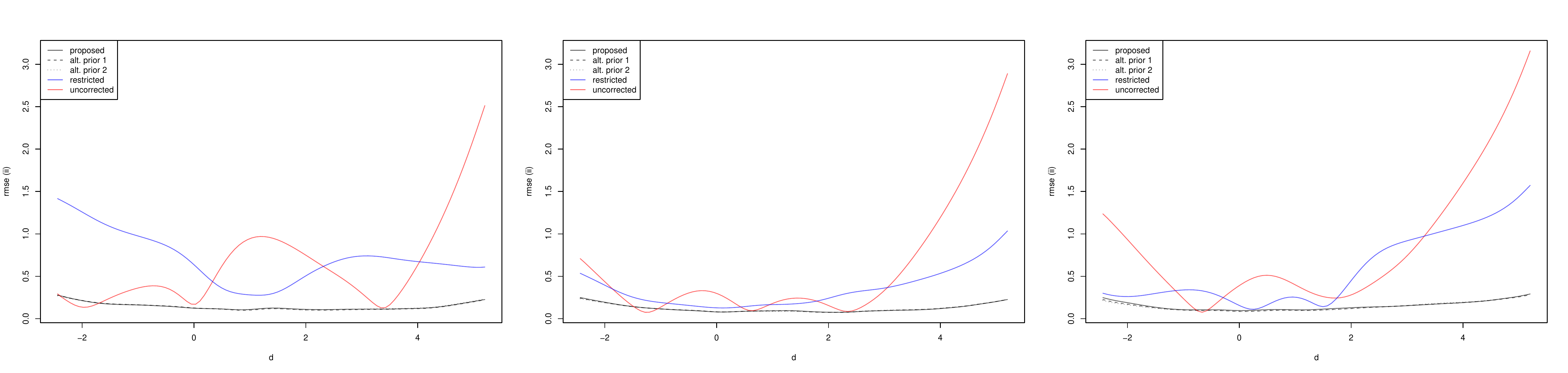}\\
\includegraphics[scale=0.2]{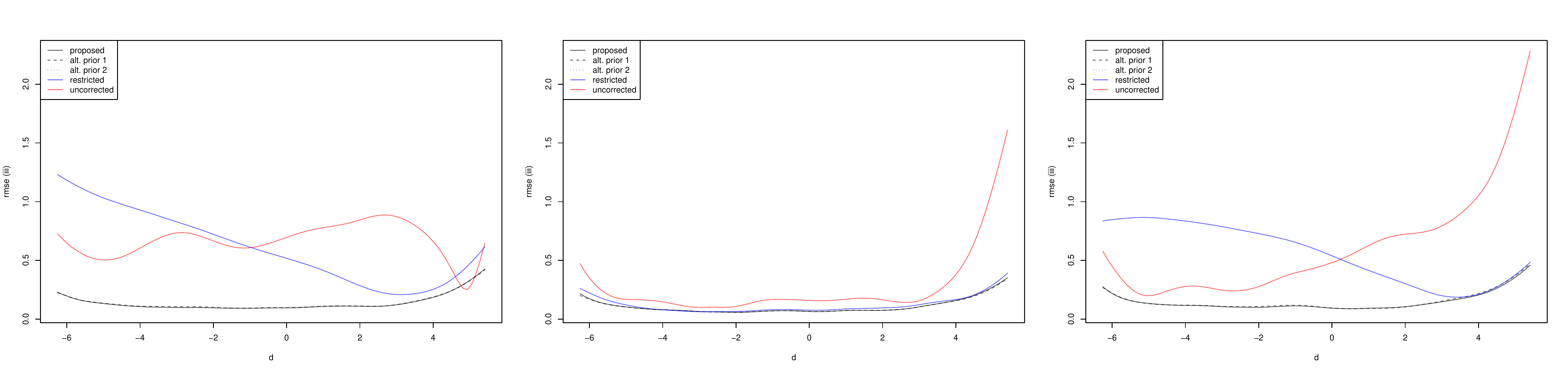}
\caption{Nonparametric models (i), (ii), and (iii): RMSEs for the proposed, restricted, and uncorrected models for $n=500$ (top three rows) and $n=2000$ (bottom three rows)}
\label{fig:sim2b}
\end{figure}

Figure~\ref{fig:sim2c} presents the estimated curves for Settings~(i), (ii), and (iii) over 100 replication in the proposed model with the default and alternative prior specifications, restricted model, and uncorrected models for $p=0.1, 0.5$, and $0.9$ for $n=2000$.
As in the previous simulation study, the figure shows that ignoring the covariate dependent variance or endogeneity leads to biased estimates.
The figure also shows that the proposed model with three prior specifications can correctly estimate the true curves and appear to provide qualitatively similar results in these cases.
Figures~\ref{fig:sim2a} and \ref{fig:sim2b} present the biases and RMSEs for $n=500$ and $n=2000$ and $p=0.1, 0.5$, and $0.9$.
Figure~\ref{fig:sim2a} shows that the restricted and uncorrected models produce biased estimates regardless of the sample size for all quantile levels and the biases nonlinearly depend on the value of $d$.
By contrast, the bias for the proposed model is near zero for all values of $d$ and becomes closer to zero as the sample size increases.
As shown in Figure~\ref{fig:sim2b}, the RMSE for the proposed model is smaller than those for the restricted and uncorrected models for most parts of the region where $d$ is observed.
Here as well, the RMSEs for the restricted model and, particularly, the uncorrected model, substantially vary by the value of $d$.
The figures also show that there is little difference in the biases and RMSEs between the default prior specification and with the two alternative prior specifications.

\begin{figure}[H]
\centering
\includegraphics[scale=0.2]{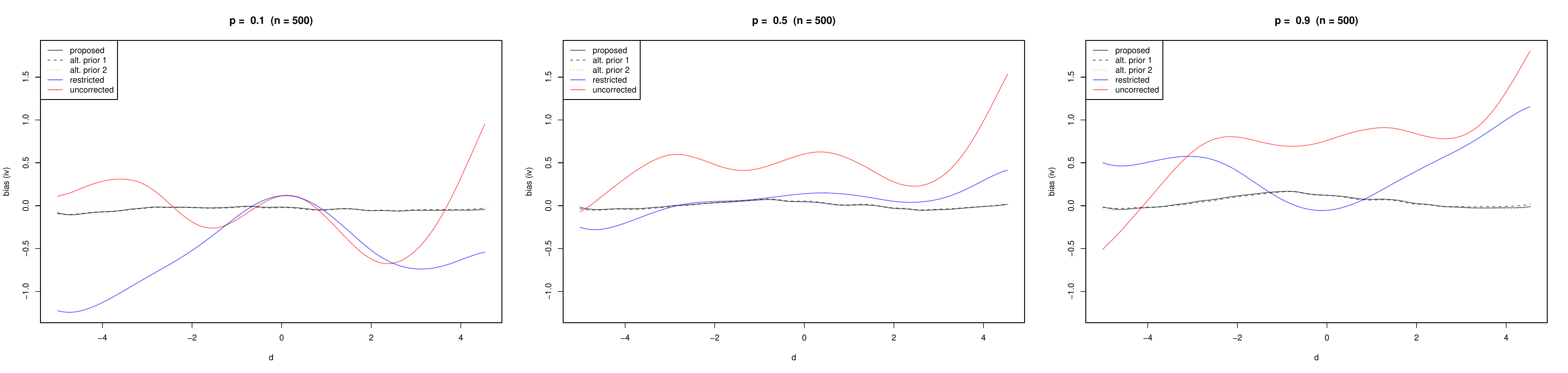}\\
\includegraphics[scale=0.2]{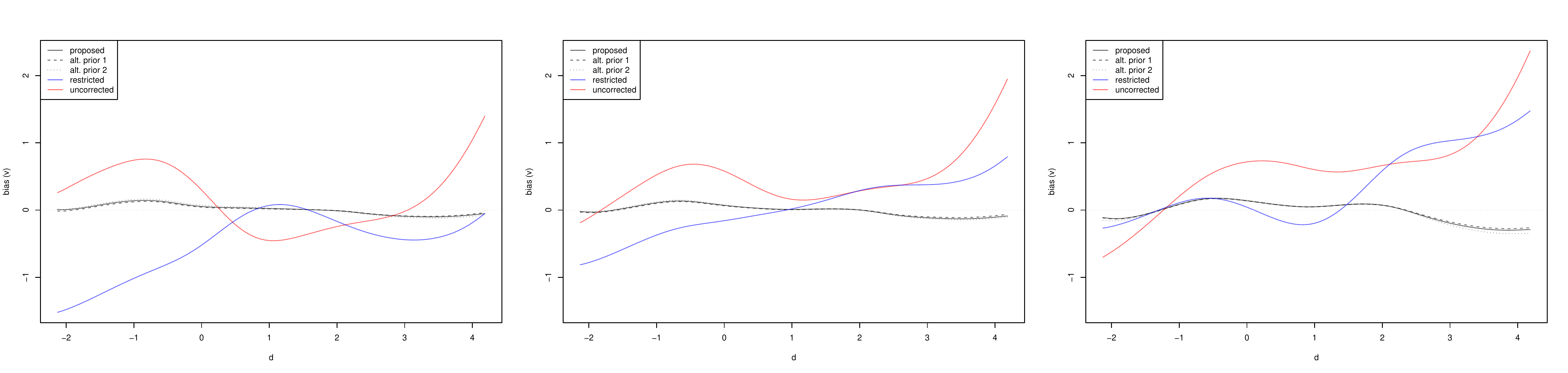}\\
\includegraphics[scale=0.2]{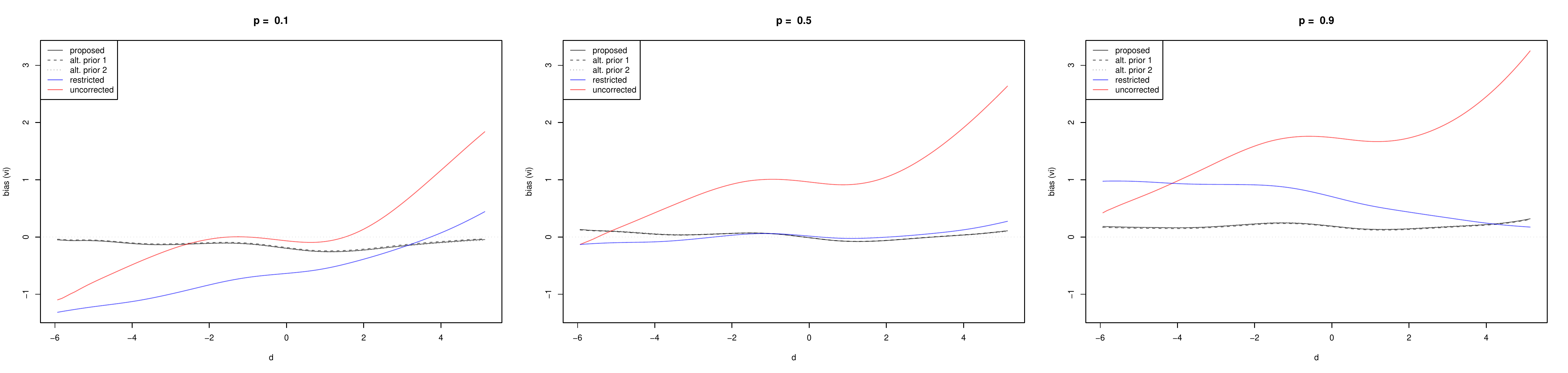}\\
\includegraphics[scale=0.2]{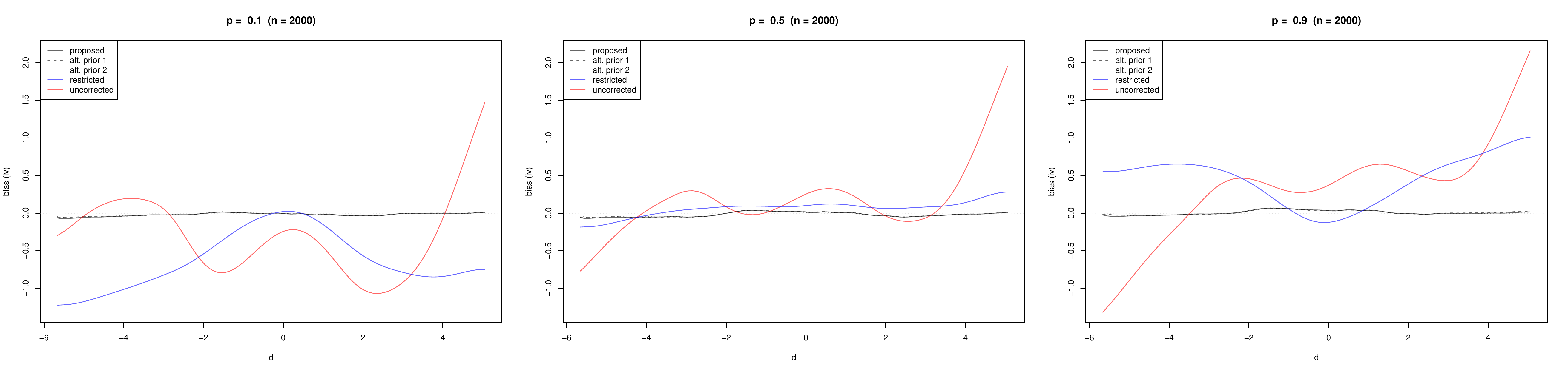}\\
\includegraphics[scale=0.2]{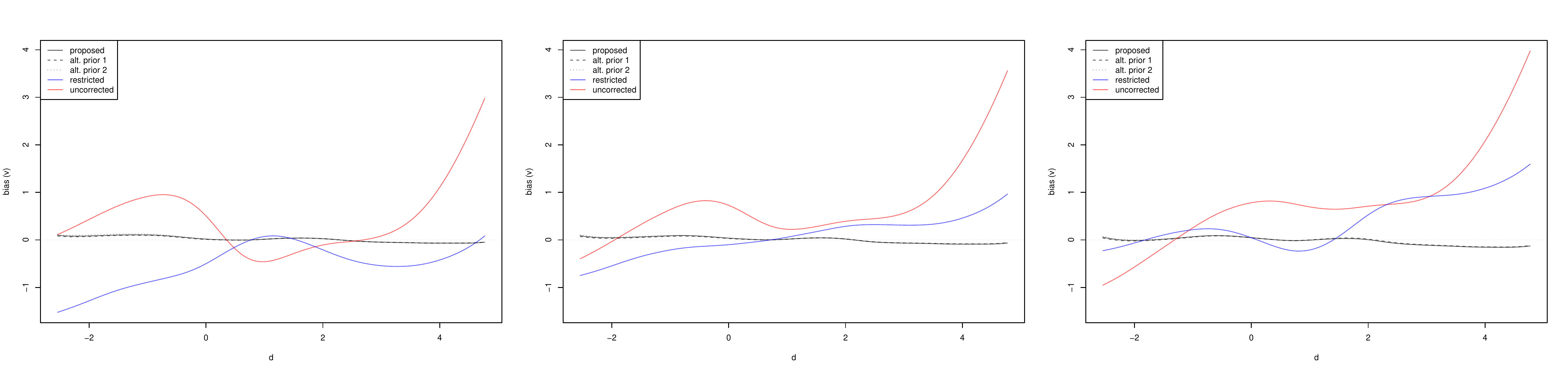}\\
\includegraphics[scale=0.2]{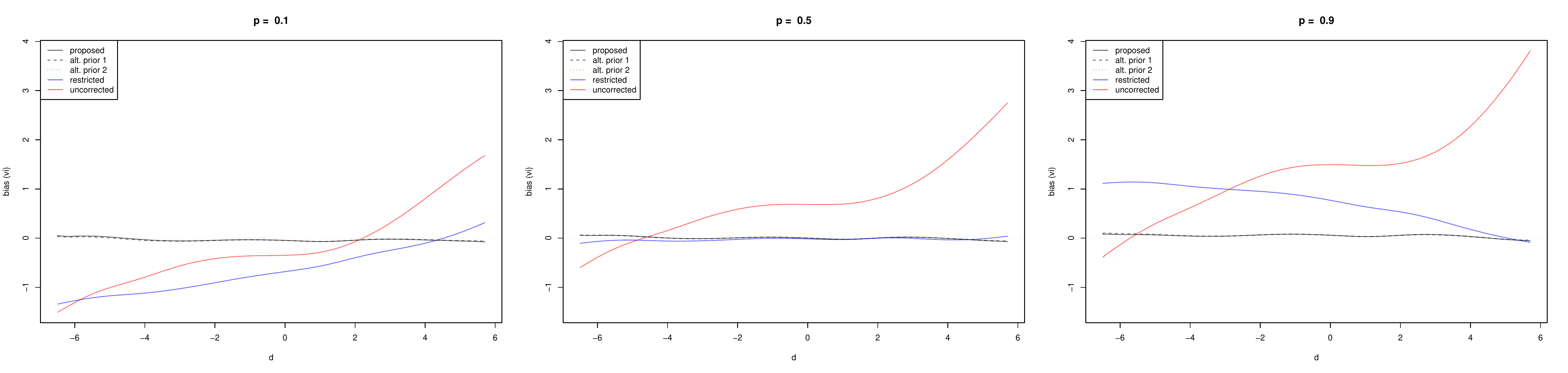}
\caption{Nonparametric models (iv), (v), and (vi): biases for the proposed, restricted, and uncorrected models for $n=500$ (top three rows) and $n=2000$ (bottom three rows)}
\label{fig:sim2d}
\end{figure}

\begin{figure}[H]
\centering
\includegraphics[scale=0.2]{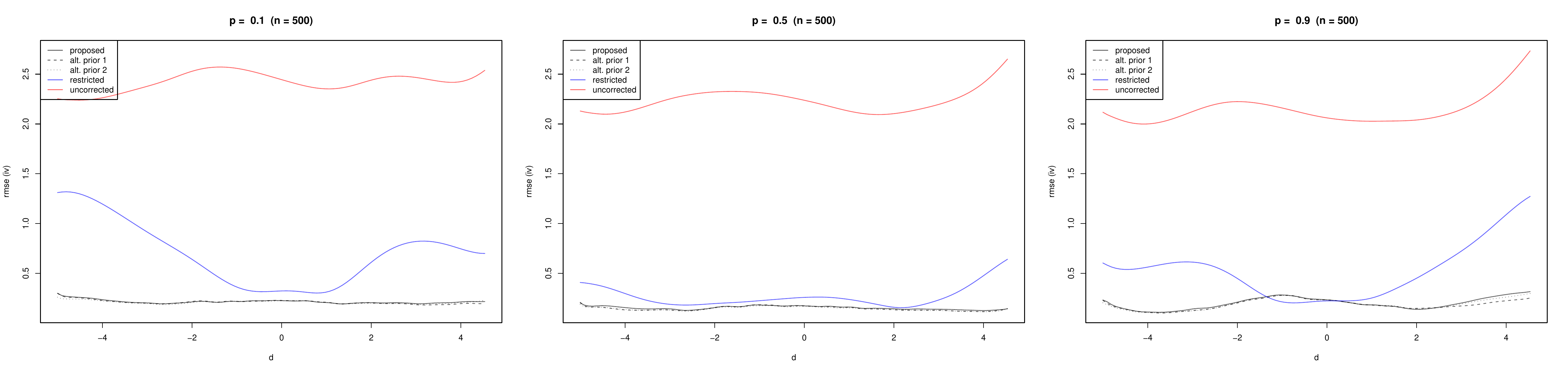}\\
\includegraphics[scale=0.2]{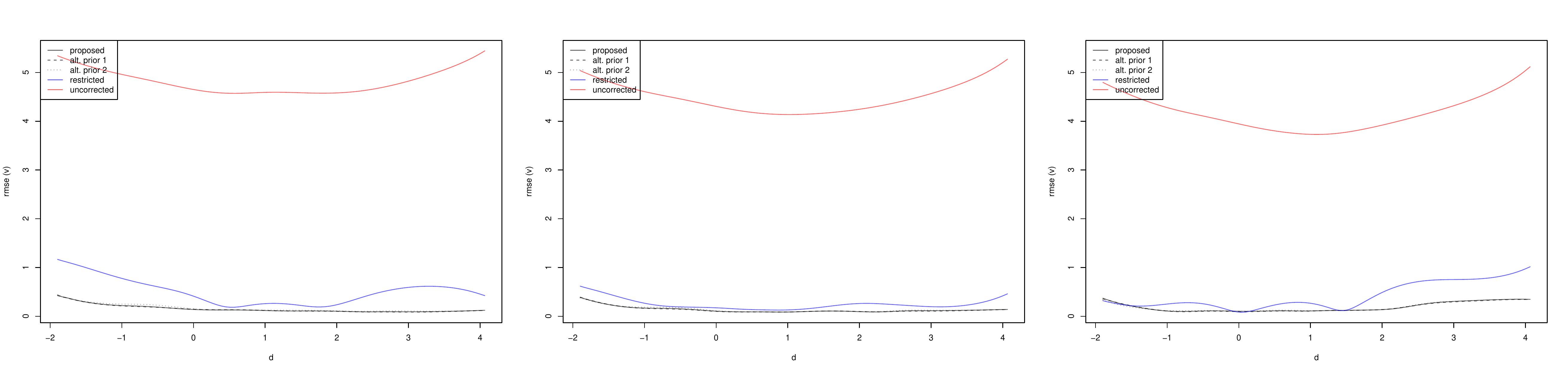}\\
\includegraphics[scale=0.2]{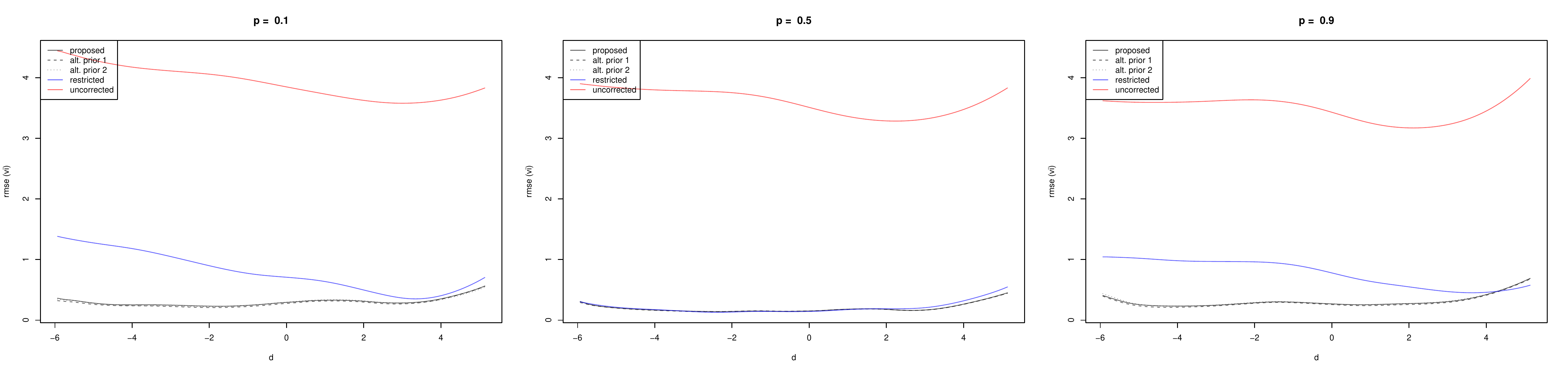}\\
\includegraphics[scale=0.2]{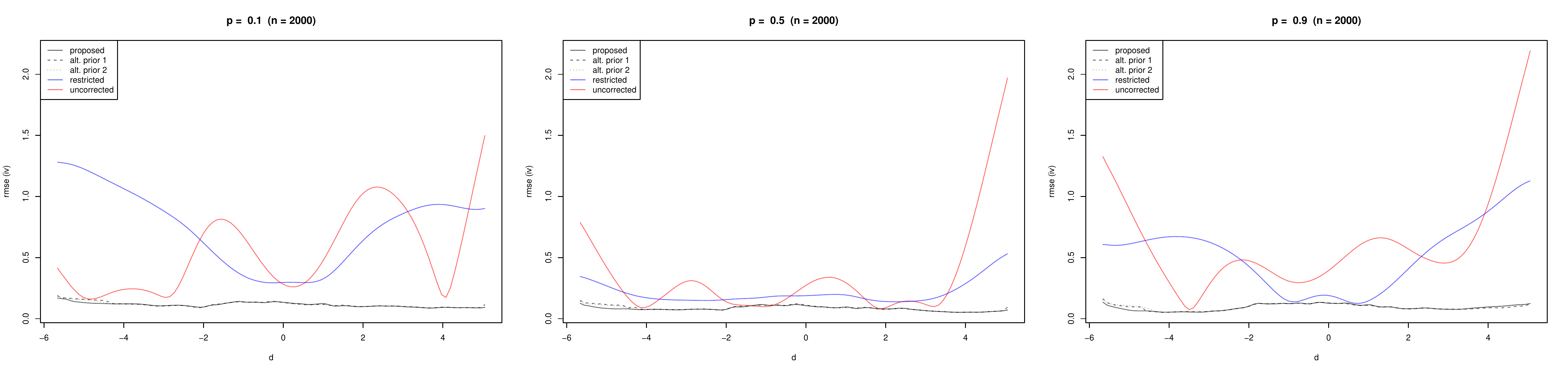}\\
\includegraphics[scale=0.2]{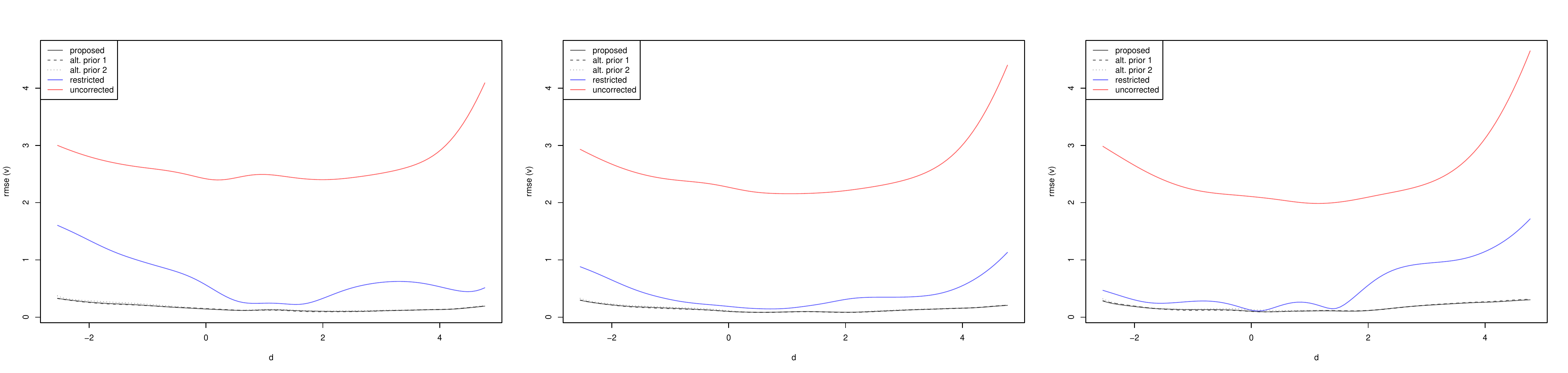}\\
\includegraphics[scale=0.2]{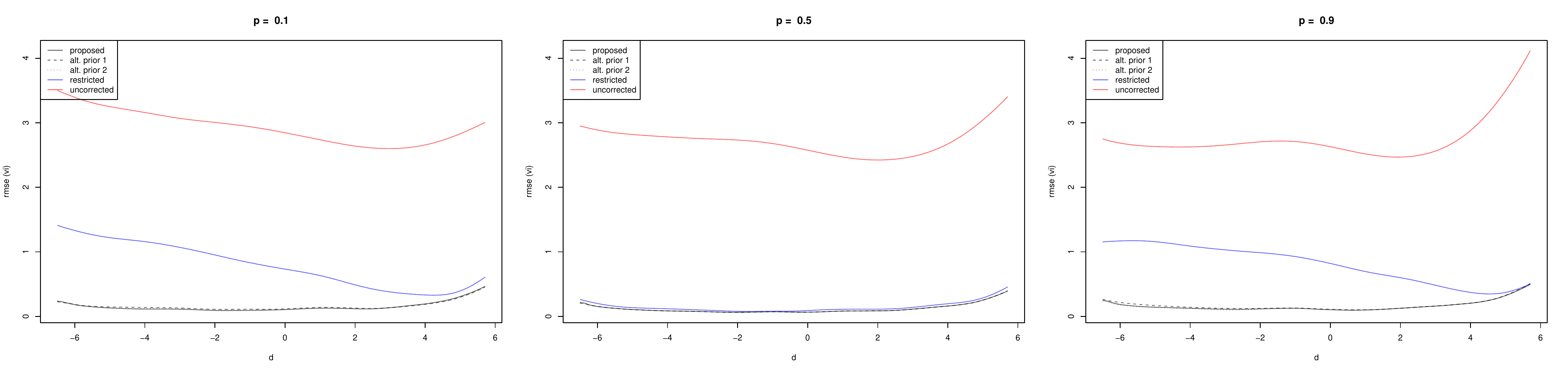}
\caption{Nonparametric models (iv), (v), and (vi): RMSEs for the proposed, restricted, and uncorrected models for $n=500$ (top three rows) and $n=2000$ (bottom three rows)}
\label{fig:sim2e}
\end{figure}

The proposed model does not model the variance of $v$ as a function of the covariates and the variances of $v$ in the data-generating models do not depend on the covariate in Settings~(i), (ii), and (iii).
Therefore, to examine how the proposed model performs when the variance of $v$ is a function of the covariate, Settings (i), (ii), and (iii), respectively, are extended to (iv), (v), and (vi), such that $v_i\sim\N(\mu_{1i},t(z_i)\tau_i)$ with the following settings
\renewcommand{\labelenumi}{(\roman{enumi})}
\begin{enumerate}
\setcounter{enumi}{3}
\item
$t(z)=\exp(-0.2z)$,
\item
$t(z)=4\N(z;0,1)$,
\item
$t(z)=4.0(\N(z;1,0.5)+\N(z;-1,0.5))$.
\end{enumerate}
Figures~\ref{fig:sim2d} and \ref{fig:sim2e} show the biases and RMSEs for Settings~(iv), (v), and (vi) under the proposed, restricted, and uncorrected models for $p=0.1$, $0.5$, and $0.9$ and for $n=500$ and $2000$.
Similar to Figure~\ref{fig:sim2a}, the bias for the proposed model is near zero for all values of $d$ and becomes closer to zero as the sample size increases.
We obtained similar results for the RMSE, which are also shown in Figure~\ref{fig:sim2b}.
Furthermore, the results for these settings under the proposed model do not appear to be significantly influenced by the prior specification.

\section{Real Data Example: Death Rates in Inter-war Japan}\label{sec:real}
\subsection{Background}
To illustrate the proposed approach using real data, the effect of the number of doctors on Japan's death rate during the inter-war period is studied.
The economic history literature has extensively studied declines in the mortality rates of Western countries across the 19th and 20th centuries.
Recent studies have found the improving effects of national health insurance systems on mortality rate (\eg\ Winegarden and Murray,~1998; 2004; Bowblis,~2010).
The successful application of germ theory has also been considered to decrease the mortality rate (Mackenbach,~1996; Mokyr and Stein,~1997).
To the effect of medical care, the importance of opportunity for medical care, for instance, through one-on-one contact and medical vouchers, for people with limited access to medical care have also been stressed in recent studies (Moehling and Thomasson,~2014; Ogasawara and Kobayashi,~2015).
These studies revealed that increased access to medical care contributed to the decline in mortality rates in the process of economic development.

While the existing literature has dominantly focused on the \textit{indirect} effects of medical doctors through national insurance systems or medical vouchers, limited works identify the \textit{direct} effects of medical doctors on the mortality rate in industrialising countries.
Evaluating the effects of medical access on mortality rate has key implications for medical policies in developing countries (\eg\ Banerjee and Deflo,~2007; Schultz,~2010).
Since the number of medical doctors in a certain area may be correlated with unobservable wealth levels, an instrumental variable technique is needed to straightforwardly identify the effects of medical doctors on mortality rate.
Therefore, we estimate the impact of medical doctors with modern medical techniques on the crude death rate (hereafter, death rate) by applying the proposed approach to the newly constructed county-level Japanese dataset for 1925 and 1930.

\subsection{Data and Model}
To analyse the effects of an increase in medical opportunities on death rate, we compiled a large set of statistical reports published by the central and local governments, such as the Prefectural Statistical Tables (PSTs) and Population Census for 1925 and 1930.
The finer details of the documents and method of constructing the dataset are provided in \ref{sec:appb}.
We restricted the sample to counties in the mainland of Japan because of the small sample size in several small islands.
The data comprise 638 observations for each year, and thus, a total of 1,276 observations are used in the following analysis.

Table~\ref{tab:real1} presents the summary statistics and the definition of the variables included in our data.
The death rate (\dr) is defined as the number of deaths per 1,000 people.
The sample means of the death rate are $21.299$ and $19.402$ for 1925 and 1930.
Our key variable is the log of the number of doctors (\lndoc).
The sample means of $\lndoc$ in our data are $3.756$ and $3.793$ for 1925 and 1930.
For the above-mentioned reason, it is treated as an endogenous variable.
The effects of the number of doctors on the mean and variance of the death rate are allowed to vary between the years by considering a varying coefficient specification.
The instrumental variable used herein is the log of the total area of the county in square kilometres (\lnarea).
We argue that our exclusion restriction is plausible because the borders of the counties were exogenously predetermined and thus, it should not be affected by unobserved factors, which might have affected medical opportunities such as potential wealth levels in the counties.
It will be confirmed that the area has a positive effect on the number of medical doctors.

The covariates include the share of infant (\child) and elder (\old) population and the age structure of the population is assumed to have nonlinear effects on the death rate.
We also include the log odds ratios of the proportion of population employed in the agricultural (\agri), mining (\mine), and industrial (\ind) sectors and the log odds ratio of the primary school enrolment rate (\enrol).
These variables are included to control for the population and industrial structure effects, potential wealth level, and unobservable macroeconomic shock in a given year and are assumed to have linear effects.
To summarise, the following model is fitted to the data:
\begin{equation}\label{eqn:emp}
\begin{split}
\lndoc&=\gamma_0+f_{1}(\child)+f_{2}(\old)+f_{3}(\lnarea)+\vx_l'\vgamma_l+\eps_1,\\
\dr&=\beta_0+g_1(\child)+g_2(\old)+g_3(\lndoc)+g_4(\lndoc)\dum +\vx_l'\vbeta_l+\eps_2,\\
s&=\exp\left\{\alpha_0+h_{1}(\textit{child})+h_2(\old)+h_3(\lndoc)+h_4(\lndoc)\dum+\vx_l'\valpha_l\right\},
\end{split}
\end{equation}
where $\vx_l=(\agri,\ind,\mine,\enrol)'$ and $\dum,\ t\in\{\ybf,\yaf\}$ is the dummy variable such that $D(\ybf)=0$ and $D(\yaf)=1$.

\begin{table}
\centering
\caption{Real data: summary of data}
\label{tab:real1}
{\scriptsize
\begin{tabular}{llrrrr}\toprule
Variable & Definition & Mean & SD & Min & Max \\\hline
\dr     & Number of deaths per 1,000 people                                                          &  20.350 & 2.570 &  11.871 & 33.412 \\
\lndoc  & Log of number of doctors                                                              &   3.775 & 0.817 &   0.693 &  8.313 \\
\agri   & Log odds ratio of proportion of population employed in agricultural sector    &  -1.225 & 1.139 &  -6.185 &  0.114 \\
\ind    & Log odds ratio of proportion of population employed in industrial sector      &  -2.574 & 0.580 &  -4.401 & -0.170 \\
\mine   & Log odds ratio of proportion of population employed in mining sector          &  -7.263 & 1.643 & -12.345 & -1.240 \\
\enrol  & Log odds ratio of primary school enrolment rate                                       &   5.592 & 0.621 &   2.674 &  9.210 \\
\child  & Proportion of infant population                                                       &   0.171 & 0.014 &   0.101 &  0.217 \\
\old    & Proportion of elder population                                                        &   0.085 & 0.019 &   0.306 &  1.443 \\
\lnarea & Log of total area in square kilometres                                                &   3.188 & 1.501 &   0.693 &  8.313 \\\bottomrule
\end{tabular}
}
\end{table}

The default prior distribution described in Section~\ref{sec:prior} and the same number of interior knots as in Section~\ref{sec:sim2} are used.
The Gibbs sampler is run for 40,000 iterations after the burn-in period of 2,000 iterations.
Every tenth Gibbs draw is retained for posterior inference.

\subsection{Results}
\begin{figure}[H]
\centering
\includegraphics[scale=0.35]{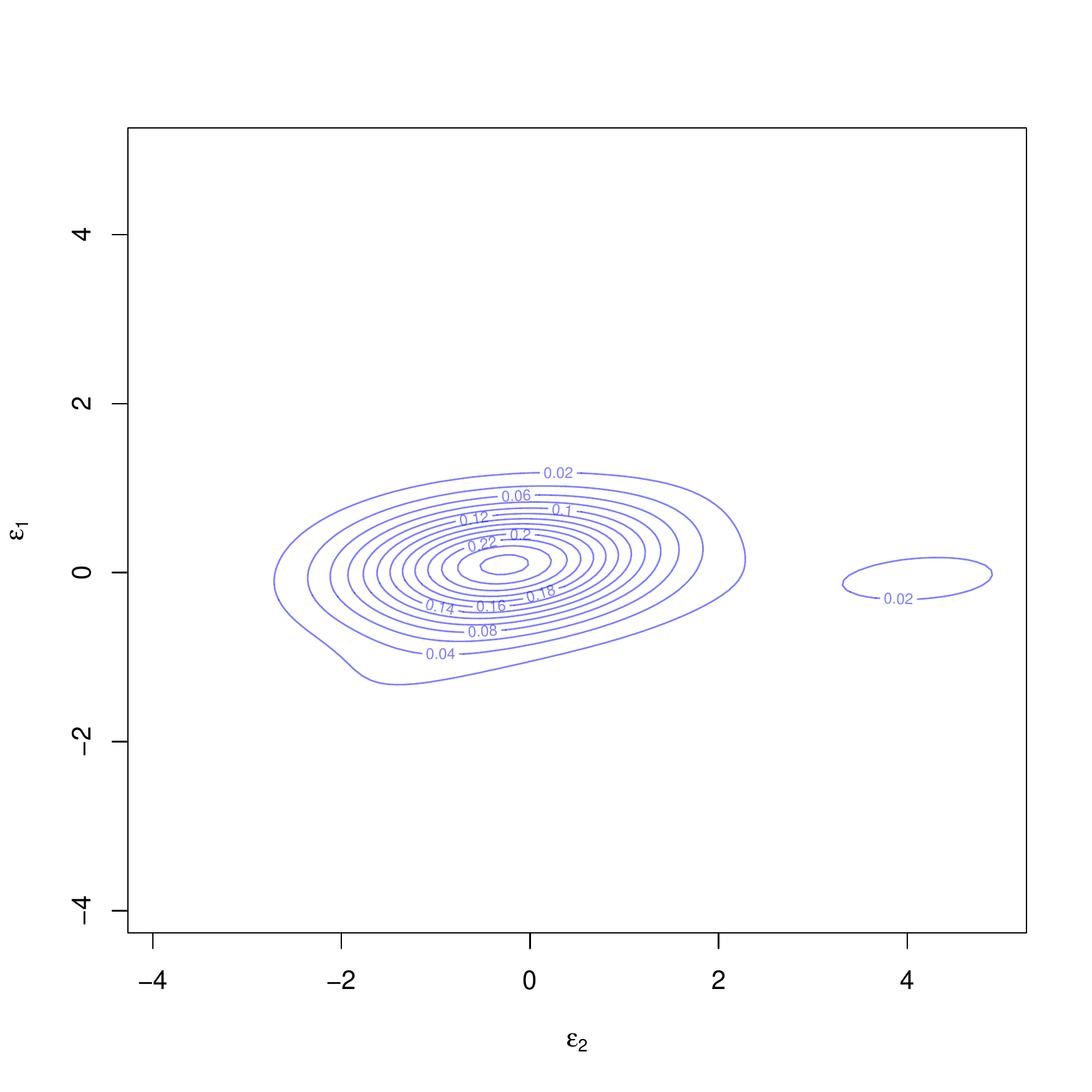}
\caption{Real data: estimated joint error density}
\label{fig:real1}
\end{figure}

\begin{figure}[h]
\centering
\includegraphics[width=\textwidth]{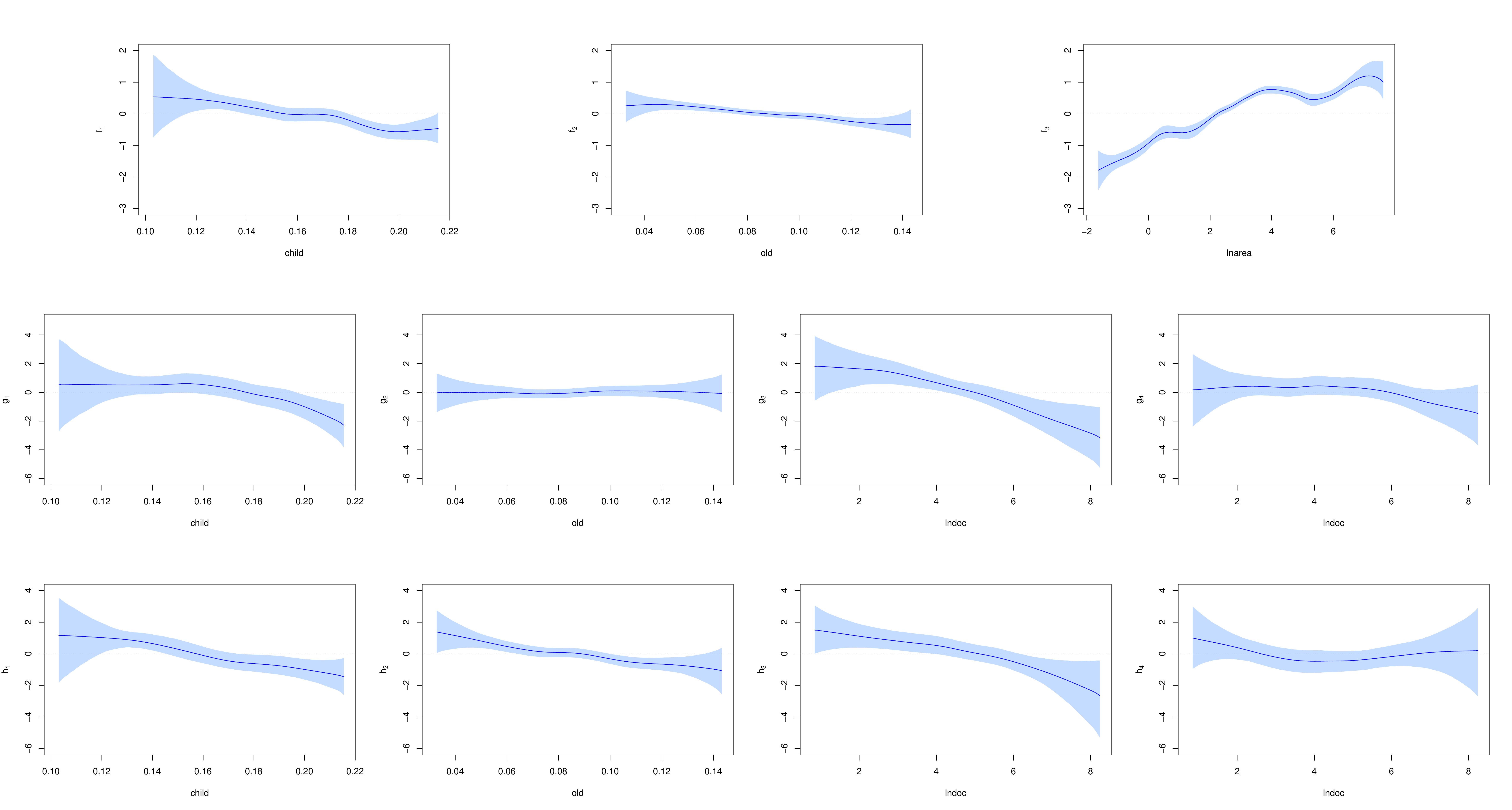}
\caption{Real data: posterior means and 95\% credible intervals for nonlinear effects in mean and variance functions}
\label{fig:real2}
\end{figure}

First, the results for error density and mean and variance functions are presented.
Figure~\ref{fig:real1} presents the estimate for the joint error density.
The figure shows that the error density exhibits a slight deviation from normality.
Figure~\ref{fig:real2} presents the posterior means and 95\% credible intervals for the nonlinear effects in the mean and variance functions.
It is shown that $\lnarea$ is able to explain the variation in $\lndoc$ and appears to be a valid instrument.
Moreover, the estimate for $f_3$ exhibits some nonlinearity.
From the estimate of $g_3$, our key variable $\lndoc$ has some nonlinearly improving effect on the mean death rate.
The estimate of $g_4$ shows that the mean death rate for the counties with a large number of doctors may be lower in 1930 than in 1925, although the 95\% credible interval includes zero for the entire range of $\lndoc$.
The conditional variance of the death rate appears to vary nonlinearly with $\lndoc$ as well.
The estimate of $h_3$ shows that the conditional distribution of the death rate has a large variance for the small values of $\lndoc$ and decreases as $\lndoc$ increases.
Table~\ref{tab:real2} presents the posterior means, 95\% credible intervals, and inefficiency factors for the linear effects.

\begin{table}[H]
\caption{Real data: posterior means, 95\% credible intervals (CI), and inefficiency factors (IF) for linear effects}
\label{tab:real2}
\centering
{\small
\begin{tabular}{lr@{\quad(}r@{,\ }r@{)\quad}rr@{\quad(}r@{,\ }r@{)\quad}rr@{\quad(}r@{,\ }r@{)\quad}r}\toprule
&\multicolumn{4}{c}{$\vgamma_l$}&\multicolumn{4}{c}{$\vbeta_l$}&\multicolumn{4}{c}{$\valpha_l$}\\
\cmidrule(lr){2-5}\cmidrule(lr){6-9}\cmidrule(lr){10-13}
Variable& \multicolumn{1}{c}{Mean} &\multicolumn{2}{c}{95\% CI}& IF &\multicolumn{1}{c}{Mean} &\multicolumn{2}{c}{95\% CI}& IF & \multicolumn{1}{c}{Mean} &\multicolumn{2}{c}{95\% CI}& IF \\\hline
Const.      &  2.929 &  2.426 &  3.426 & 1.0 & 16.659 &  14.969 & 18.352 & 2.0 &  0.931 & -0.699 &  2.638 &  5.0 \\
\agri       & -0.623 & -0.701 & -0.542 & 1.1 & -0.080 &  -0.339 &  0.175 & 1.2 &  0.061 & -0.130 &  0.250 &  4.3 \\
\ind        &  0.008 & -0.066 &  0.080 & 2.0 & -0.442 &  -0.711 & -0.171 & 3.1 & -0.339 & -0.595 & -0.078 &  5.1 \\
\mine       & -0.028 & -0.048 & -0.009 & 2.4 & -0.071 &  -0.145 &  0.004 & 2.9 & -0.041 & -0.104 &  0.027 &  7.0 \\
\enrol      & -0.063 & -0.115 & -0.012 & 0.9 & -0.063 &  -0.260 &  0.134 & 2.1 & -0.162 & -0.360 &  0.030 &  6.3 \\\bottomrule
\end{tabular}
}
\end{table}

Next, the results regarding the quantiles are presented.
We first study how the conditional quantiles of the death rate vary by the number of doctors.
To compare the differences in the effects of $\lndoc$ between the two years, we estimate the conditional quantiles over a grid of $\lndoc$ with either $t=\ybf$ or $\yaf$ and the remainder of the covariates fixed at the sample average in 1930.
Namely, $S_{\dr|\lndoc,\bar{\vx}_\yaf,D(t)}(p)$ is estimated for $t\in\{\ybf,\yaf\}$, where $\bar{\vx}_\yaf$ denotes the sample average for the rest of the covariate for 1930.

\begin{figure}[H]
\centering
\includegraphics[width=\textwidth]{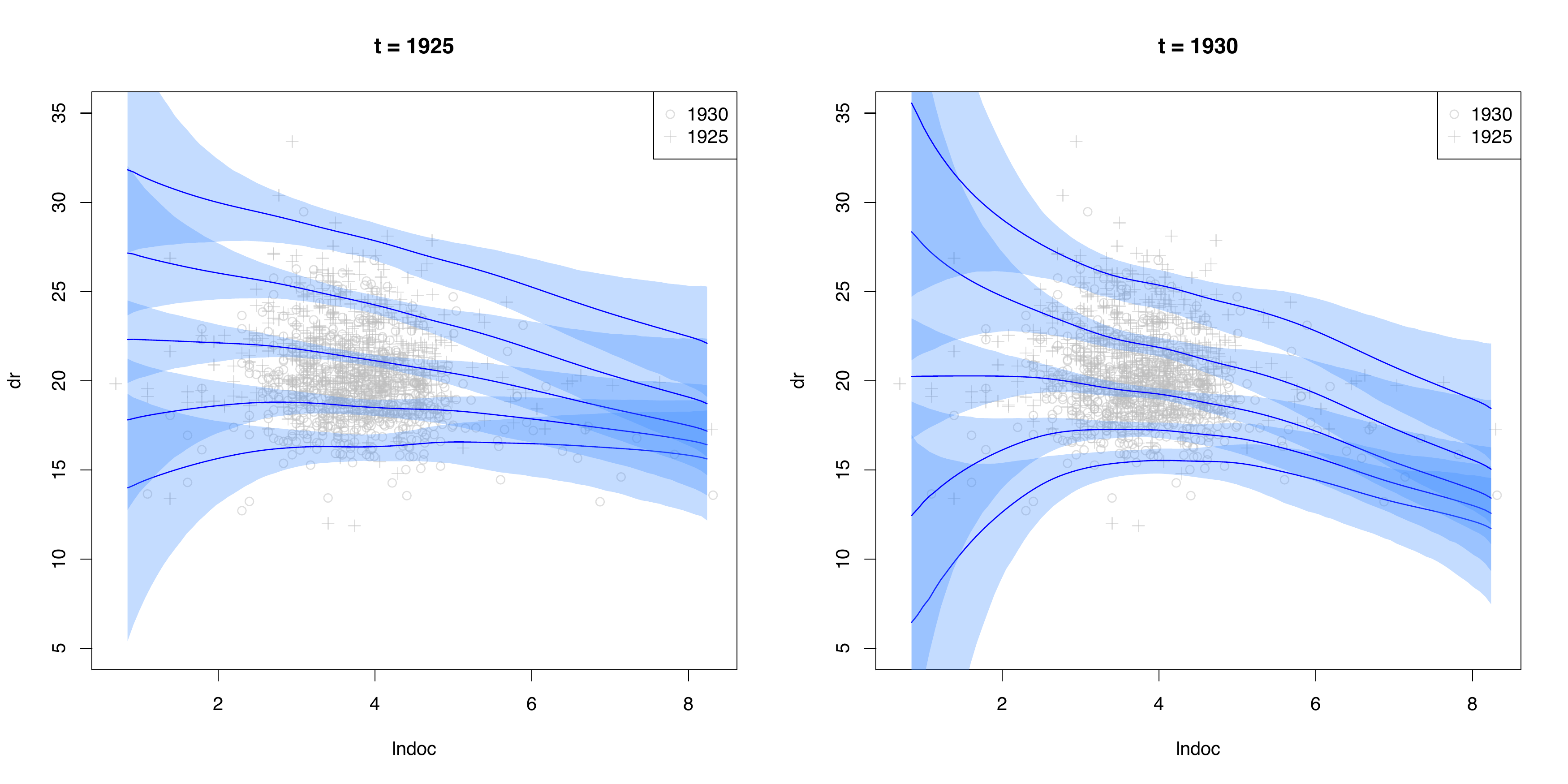}
\caption{Real data: posterior means and 95\% credible intervals of $S_{\dr|\lndoc,\bar{\vx}_\yaf,D(t)}(p)$ for
for $p=0.01, 0.1, 0.5, 0.9, 0.99$}
\label{fig:real3}
\end{figure}

Figure~\ref{fig:real3} presents the posterior means and 95\% credible intervals of $S_{\dr|\lndoc,\bar{\vx}_\yaf,D(t)}(p)$ for $p=0.01,0.1,0.5,0.9$, and $0.99$.
As the descriptive statistics suggest, the conditional quantiles for $t=\yaf$ are smaller than the corresponding quantiles for $t=\ybf$.
The figure shows that in the case of $t=\yaf$, the conditional distribution in the region, say $\lndoc<2$, is quite dispersed compared to the case of $t=\ybf$, because the conditional variance of the death rate in this region is estimated to be large, as shown in Figure~\ref{fig:real2}.
The posterior distributions of the quantiles are also dispersed. 
This can be attributed to the fact that our data include few observations in this region during 1930 (see Figure~\ref{fig:real3}).
Therefore, we limit our focus to region $\lndoc>2$.

As for the effect of doctors, the figure clearly shows that the quantiles for $p=0.5, 0.9$, and $0.99$ decrease as the number of doctors increases for both $t=\ybf$ and $\yaf$.
While the quantiles for $p=0.01$ and $0.1$ do not appear to change for the moderate values of $\lndoc$ in both cases, these quantiles start to decline for $\lndoc>5$ for $t=\yaf$.
Furthermore, for $t=\yaf$, all quantiles move downwards more quickly in this region than for the moderate values of $\lndoc$.
By contrast, for $t=\ybf$, little decline for $p=0.01$ and $0.1$ is observed in this region.

\begin{figure}[H]
\centering
\includegraphics[width=\textwidth]{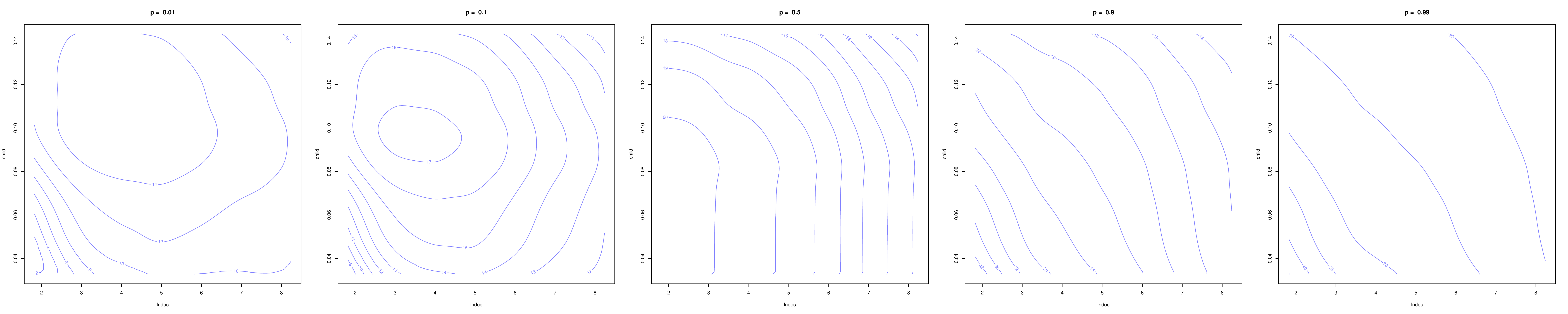}\\
\includegraphics[width=\textwidth]{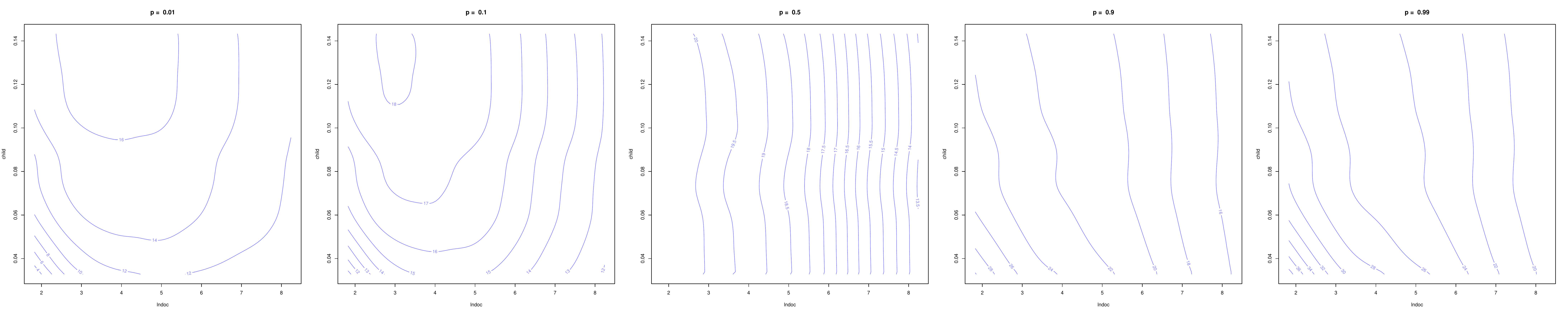}
\caption{Real data: posterior means for the joint quantile surfaces for $p=0.01$, $0.1$, $0.5$, $0.9$, and $0.99$ for $t=1930$}
\label{fig:real4}
\end{figure}

We also estimate the joint quantile surfaces over $\child$ and $\lndoc$ denoted by $S_{\dr|\child,\lndoc,\bar{\vx}_t,D(t)}(p)$ and over $\old$ and $\lndoc$ denoted by $S_{\dr|\old, \lndoc,\bar{\vx}_t,D(t)}(p)$.
Figure~\ref{fig:real4} presents the posterior means of the joint quantile surfaces for $p=0.01,0.1,0.5,0.9$, and $0.99$ for $t=\text{1930}$, showing how the quantile varies by the number of doctors and proportion of young and elder population.
It is observed that the quantiles of the death rate for $p=0.5, 0.9$, and $0.99$ decrease as the number of doctors increases, irrespective of the proportions of young and elder population.
Similar to the above result, the quantiles for $p=0.01$ and $0.1$ begin to decline for, say, $\lndoc>5$.

To summarise the results, we confirm that the death rate declines as the number of doctors increases.
The improving effect on the death rate is observed in the middle and upper quantiles for the county with a moderate number of doctors.
In particular, in 1930, the middle and upper quantiles of the death rate are found to decline at a faster rate for the county with a large number of doctors.
For such a county, even lower quantiles of death rate begin to decline as the number of doctors increases.
This result suggests that the early enforcement of the government-supported health-insurance system in 1927 accelerated the mitigating effects of doctors on the mortality rate latest by 1930.
Our finding is in line with some existing empirical results in economic history, which argues the contribution of early health insurance to the historical decline in death rate (\eg\ Winegarden and Murray~1998; 2004; Bowblis~2010).

\section{Conclusion}\label{sec:conc}
In this study, we considered a Bayesian nonparametric approach to inference for the conditional quantile of the response variable in the presence of the endogenous variable.
The proposed location-scale model maintains the familiar form of the Bayesian instrumental variable regression model and allows for a great deal of flexibility in the estimated quantile curve through the conditional variance of the second error, which varies smoothly by covariate.
The application of the varying coefficient specification to the data on Japan's death rate in 1925 and 1930 revealed that the middle and upper quantiles of the death rate decrease as the number of doctors increases.
Moreover, in 1930, for the county with a large number of doctors, the middle and upper quantiles decline at a faster rate and the lower quantiles also decline as the number of doctors increases.

\subsubsection*{Acknowledgements}
This study was supported by JSPS KAKENHI [Grant Numbers 15K17036 and 16K17153].
The computational results were obtained using Ox version 6.21 (Doornik,~2007).

\appendix
\def\thesection{Appendix~\Alph{section}}
\def\thesubsection{\Alph{section}.\arabic{subsection}}
\section{Parameters for Mixture Sampler}\label{sec:appa}
In the mixture sampler to sample $\valpha$, the log-$\chi^2$ distribution is approximated using the ten-component mixture of normal distributions:
\begin{equation*}
f(e^*)\approx\sum_{h=1}^{10}\varrho_h\N(e^*;\xi_h,\zeta_h).
\end{equation*}
The parameters for the mixture are presented in the table below.

\begin{center}
\begin{tabular}{crrr}\toprule
$h$&\multicolumn{1}{c}{$\varrho_h$}&\multicolumn{1}{c}{$\xi_h$}&\multicolumn{1}{c}{$\zeta_h$}\\\hline
1&0.00609&1.92677&0.11265\\
2&0.04775&1.34744&0.17788\\
3&0.13057&0.73504&0.26768\\
4&0.20674&0.02266&0.40611\\
5&0.22715&-0.85173&0.62699\\
6&0.18842&-1.97278&0.98583\\
7&0.12047&-3.46788&1.57469\\
8&0.05591&-5.55246&2.54498\\
9&0.01575&-8.68384&4.16591\\
10&0.00115&-14.65000&7.33342\\\bottomrule
\end{tabular}
\end{center}

\section{Data Appendix}\label{sec:appb}
The crude death rate is defined as the number of deaths per 1,000 people.
The number of deaths in each county is taken from a compilation of \textit{Fuken tokei hyo} (Prefectural Statistical Tables; PSTs) for 1925 and 1930 published by each prefecture (Yushodo Filmshuppan Yugenkaisha, 1977).
The PSTs for some prefectures for a specific year generally comprise several volumes of reports, such as the volumes for the land and population, education, public health, and police.
Each volume comprises roughly 200--1,000 pages.
In total, we compiled roughly 200 volumes to obtain the number of deaths in each county.
The population are taken from \textit{Kokusei chosa hokoku} (Population Census of Japan; PCJ) for 1925 and 1930 (Statistics Bureau of the Cabinet 1926; 1935).

Data on the number of medical doctors in each county are from the PSTs.
To account for the missing values in Miyagi Prefecture for 1925, we used data from \textit{Dainihon ishimeibo} (List of medical doctors in Japan; LMDJ 1925) edited by medical doctor Masao Takasaki.
For the missing values in Aomori, Miyagi, Fukushima, Tochigi, Shizuoka, Shimane, Hiroshima, and Fukuoka prefectures for 1930, we adopted data from \textit{Dainihon ishimeibo} (LMDJ 1930) edited by Kaneharashoten.
Data on the total area in square kilometres were also taken from the PSTs.
For the missing values in Fukui (1925 and 1930) and Miyagi and Ishikawa (in 1930) prefectures, we used the 1921 and 1925 values of total area for these prefectures from the PSTs since the total area of a prefecture is unlikely to change.

The county-level age-specific population data are from the 1925 and 1930 editions of \textit{Kokusei chosa hokoku, fukenhen} (Population Census of Japan, Prefectural Part; PCJPP) published by the Statistics Bureau of the Cabinet (various years).
For each census year, we compiled 47 reports of the PCJPP.
Data on the number of workers in each industry are also from the 1920 and 1930 editions of the PCJPP published by the Statistics Bureau of the Cabinet (various years).
Since the population census of 1925 was a partial census, the number of workers in the year is linearly interpolated using both the 1920 and 1930 values.
The data on the primary school enrolment rate are taken from the PSTs.

\end{document}